\documentclass[aps,PRL,reprint,showpacs,superscriptaddress,
footinbib,citeautoscript]{revtex4-1}

\usepackage{graphicx}

\usepackage[utf8]{inputenc}
\usepackage[export]{adjustbox}
\usepackage{csquotes}
\usepackage[american]{babel}
\usepackage[T1]{fontenc}
\usepackage{enumerate}
\usepackage{mdwlist}
\usepackage[activate=normal]{pdfcprot}
\usepackage{bbding}
\usepackage{color}
\usepackage{amsmath}
\usepackage{eqnarray,amsmath}
\usepackage{amssymb}
\usepackage{amsmath}
\usepackage{amsfonts}
\usepackage{mathrsfs}
\usepackage[titletoc,title]{appendix}
\usepackage{bm}
\usepackage{dcolumn}
\usepackage{color}
\usepackage[normalem]{ulem}
\usepackage[colorlinks=true,citecolor=blue]{hyperref}
\hypersetup{colorlinks=true,citecolor=blue,linkcolor=red
,urlcolor=blue}

\newcommand{\beq}{\begin{equation}}
\newcommand{\eeq}{\end{equation}}
\newcommand{\beqa}{\begin{eqnarray}}
\newcommand{\eeqa}{\end{eqnarray}}

\begin{document}

\title{Quadrupolar photovoltaic effect in the terahertz range in a two-dimensional spin-3/2 hole system}
\author{Mohsen Farokhnezhad}
\affiliation{School of Nano Science, Institute for Research in Fundamental Sciences (IPM), Tehran 19395-5531, Iran}
\author{W.~A.~Coish}
\affiliation{Department of Physics, McGill University, 3600 rue University, Montreal, Qc H3A 2T8, Canada}
\author{Reza Asgari}
\affiliation{School of Physics, The University of New South Wales, Sydney 2052, Australia}
\affiliation{School of Physics, Institute for Research in Fundamental Sciences (IPM), Tehran 19395-5531, Iran}
\author{Dimitrie Culcer}
\affiliation{School of Physics, The University of New South Wales, Sydney 2052, Australia}
\affiliation{Australian Research Council Centre of Excellence in Low-Energy Electronics Technologies}
\begin{abstract}
We identify a strong photovoltaic response due to the non-linear optical transition between heavy and light hole sub-bands enabled by $T_d$-symmetry in a quantum well, which we term the \textit{quadrupolar photovoltaic effect} (QPE). The photovoltaic current exhibits a strong resonance in the vicinity of the heavy hole-light hole splitting, with a magnitude governed by the momentum relaxation time, which can reach nanoseconds in GaAs holes. Since the heavy hole-light hole splitting can be tuned from a few meV to nearly 100 meV the QPE could serve as the basis for a terahertz photo-detector. We discuss strategies for experimental observation and device applications.
\end{abstract}
\date{\today}
\maketitle

\textit{Introduction}-The past decade has witnessed a spectacular resurgence in the study of non-linear electromagnetic effects, motivated by the rise of topological materials and cutting-edge developments in semiconductors~\cite{Boyd2007,papadopoulos2006non,green2006third,morimoto2016topological,PhysRevX.11.011001,PhysRevB.95.035134,PhysRevB.94.035117,PhysRevB.91.125424,culcer2020transport,PhysRevB.102.085202,asgari2021}. Second-order responses require inversion symmetry breaking and this is satisfied by most topological materials, which has led to discoveries such as Hall effects~\cite{Du_NC2019,Nandy_PRB2019,PhysRevLett.115.216806} in time-reversal invariant systems, and advances in generating non-reciprocal currents~\cite{Nagaosa_NRR_NC2018, PhysRevLett.122.227402, tzuang2014non, shao2020non}. Among the latter, photo-currents are intimately related to the Hilbert space topology and underlie photovoltaic devices ~\cite{PhysRevX.11.011001, belinicher1980photogalvanic,belinicher1978space, Ivchenko_SovPhy1984,Khurgin_JOS1994,PhysRevB.23.5590, fridkin2001bulk, PhysRevLett.119.067402, Nature493, PhysRevLett.7.118,PhysRevB.100.064301,osterhoudt2019colossal,nakamura2017shift,zhang2019enhanced, carvalho2019nonlinear, PhysRevB.61.5337, yang2010above, wang20182d, PhysRevB.79.081406}, with potential applications in solar cells, energy harvesting and terahertz devices ~\cite{green2020tracking, rogalski2000infrared, rogalski2020hgcdte, yang2019monolayer, lei2015progress, rogalski2005hgcdte}. 


A group of materials that lacks inversion symmetry is that of zinc-blende semiconductors. Recent years have witnessed a flurry of interest in holes in III-V zinc-blende semiconductors such as GaAs, which have a spin-3/2, enabling physics that is impossible in spin-1/2 electron systems. Hole systems have been synthesised to high quality exhibiting very large mobilities, display strong topological effects \cite{manfra2005high,PhysRevLett.106.236601,habib2009spin,PhysRevLett.118.146801,PhysRevLett.85.4574,PhysRevLett.113.046801,PhysRevLett.126.256601,PhysRevLett.121.087701,PhysRevB.101.121302,PhysRevMaterials.6.034005,PhysRevX.8.021068}
and are intensively studied for all-electrical quantum computing applications~\cite{chatterjee2021semiconductor,veldhorst2015two,hendrickx2020fast,PhysRevResearch.3.013081}. Until recently inversion-breaking tetrahedral-symmetry terms were believed to be negligible for holes~\cite{winkler2003spin}. Hence photovoltaic effects, which require inversion symmetry breaking, have not been investigated in purely hole systems. Yet recent research ~\cite{PhysRevB.102.075310} has revealed that tetrahedral-symmetry terms can be large, and the combination of spin 3/2 and tetrahedral symmetry results in a quadrupole spin-orbit interaction with electric fields. This interations opens the door to photovoltaic applications in the elusive terahertz range. 

\begin{figure}[t]
\includegraphics[scale=0.5]{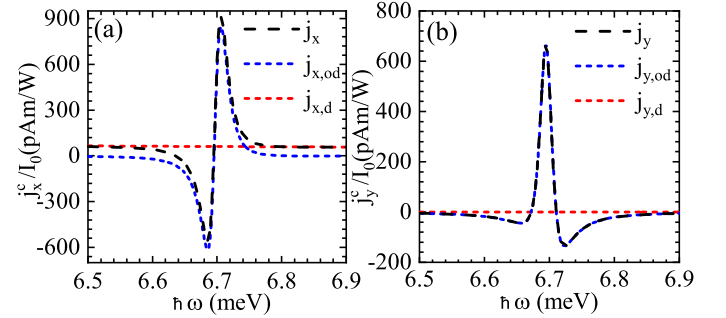}
\caption{The optical current $j$ along the directions (a) $x$ and (b) $y$. The contributions $j_{d}$ and $j_{od}$ come from the diagonal and off-diagonal parts of the density matrix, respectively. The QPE peaks occur at $\hbar{\omega} = \varepsilon^{+}_{k_F} - \varepsilon^{-}_{k_F}$ and $\hbar{\omega} = \varepsilon^{+}_{k_F} - \varepsilon^{-}_{k_F} - \delta\omega$ where $\delta \omega = \sqrt{3/4}(\frac{\hbar}{\tau_{2}})^2/(\varepsilon^{+}_{k_F}-\varepsilon^{-}_{k_F})$. The peak of the nonlinear anomalous Hall current occurs at the optical transition $\hbar{\omega}=\varepsilon^{+}_{k_F}-\varepsilon^{-}_{k_F}$ and is smaller than the main peak along $x$. Here $F_z=1\,MV/m$, $\tau_{1}=\tau_{2}=\langle \tau \rangle= 39$ ps (see Eq. (S37)), $U_0=2.5$ eV m$^{-1}$, $n_i=2.3\times{10^{11}}$ cm$^{-2}$, and other parameters are given in Table I in SM III.  The Fermi energy $\varepsilon(k=k_{F})=20.9$ meV and $k_F=0.1$ nm$^{-1}$ for the LH band; $\varepsilon_{k_{F}} = \varepsilon_{0} - \sqrt{(\frac{\Delta{\epsilon}}{2})^2 + \left({\lambda}'k_{F} + {\gamma}'_{1}k_{F}^3 + {\gamma}'_{2}k_{F}^5 + {\gamma}_{R3}k_{F}^7\right)^2}
$. The HH and LH band extrema are $10.4$ and $17.1$ meV, respectively.}
\label{Fig4}
\end{figure}

In this work, we determine the full photovoltaic response of a doped asymmetric hole GaAs quantum well, and identify a strong resonance due to optical transitions between the lowest LH and HH sub-bands. We refer to this as the \textit{quadrupolar photovoltaic effect} (QPE). The transition is caused by tetrahedral symmetry terms that go beyond the Luttinger Hamiltonian, and are responsible for an asymmetry in transition rates across the Fermi surface. The effect relies on finite doping and disorder and is not captured by a naive application of Fermi's Golden Rule. Our central result is summarized in Fig.~\ref{Fig4} where the QPE current along the $x$ and $y$ directions is plotted. The longitudinal current ($\hat{x}$ direction) is accompanied by a smaller non-linear anomalous Hall current ($\hat{y}$ direction). Since the energies involved lie in the terahertz range and the hole band gap of GaAs can be adjusted by the top gate field, this effect could be used to design a terahertz radiation detection device~\cite{rogalski2000infrared, rogalski2020hgcdte}. Hole mobilities can be orders of magnitude larger than the conservative estimate used in Fig.~\ref{Fig4} \cite{davies1991growth,kane1993variable,simmons1997fabrication,dobbie2012ultra,sigle2021strained}, leading to much larger peaks.




\textit{Model and Theory} - The hole dispersion relation is determined by the strong spin-orbit interaction in the Luttinger Hamiltonian~\cite{winkler2003spin}. In a quantum well (QW), size quantization breaks the four-fold degeneracy of the $J=\frac{3}{2}$ states and we therefore have HH states with $m_z=\pm \frac{3} {2}$ and LH states with $m_z= \pm \frac{1}{2}$. The band Hamiltonian ${\cal{H}}_{0}$ for the QW is defined as
${\cal{H}}_{0}={\cal{H}}_L + V(z)I_4 + {\cal{H}}_{d_z}$
where $I_4$ is the $4\times 4$ identity matrix and the Luttinger Hamiltonian within the spherical approximation is ${\cal{H}}_L=\frac{\hbar^2}{2m}\left[\left(\gamma_1+\frac{5}{2}\gamma_2\right)k^2 I_4-2 \gamma_2(\bm{k}\cdot\mathbf{J})^2\right]$ with parameters $\gamma_1=6.85$ and $\gamma_2=$2.10. We consider an asymmetric QW, growth along the $z$ direction, and Cartesian coordinates are aligned with the main crystal symmetry axis, formed at a hetero-interface and its confinement can be described by a triangular potential $V(z)= -e F_z z$ for $z>0$ and $\infty$ otherwise, where $e=-\vert{e}\vert$ is the electron charge and  $\textbf{F}=F_{z}\hat{z}$ is an electric field. 
In addition, the effective projected electric-dipole Hamiltonian ${\cal{H}}_{d_z}$ induced by the electrical field is expressed as \cite{PhysRevB.102.075310} $
 H_{d_z}={\frac{1}{\sqrt{3}}}e{a_B}{\chi}{F_z}{\lbrace{J_{x},J_{y}}\rbrace}
$ with $e{a_{B}}\simeq2.5$D ($a_{B}$ is the Bohr radius and $D$ is a Debye), $\chi$ is a parameter that controls the strength of the electric-dipole matrix elements. This term couples the 3/2 and -1/2 states, as well as -3/2 to 1/2, allowing for HH-LH transitions that would be otherwise forbidden. Note that there are two separate electric fields in the system: the static gate electric field causes the quantum well to be inversion asymmetric, while the applied oscillating electric field optically excites carriers. Each electric field generates its associated electric dipole term, both of which play essential roles in the QPE. The dipole term due to the gate affects the QW energy dispersion, while the one due to the applied field leads to an additional driving term, which will be discussed below. These fields, together with $T_d$ symmetry, set the direction of the second-order DC response: $T_d$ symmetry implies that, for example, $\hat{\bm x}$ is not equivalent to $-\hat{\bm x}$, while the orientation of the external electric fields determine the direction of the current as $\hat{\bm x}$ or $-\hat{\bm x}$.

\begin{figure}
\includegraphics[scale=0.6]{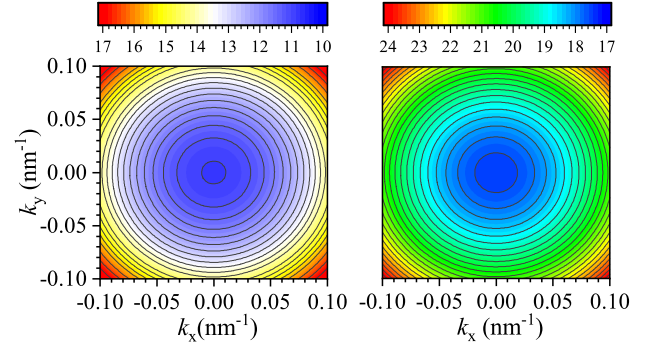}
\caption{Contour plot of the energy dispersion of the light hole and heavy hole bands around $k=0$ for a nonzero electric dipole matrix element $\chi$.}
\label{Fig.1}
\end{figure}

Using second-order degenerate perturbation theory \cite{PhysRevB.102.075310} we project the Hamiltonian matrix onto the hole subspace; see SM II and III. This gives the following effective $2 \times 2$ Hamiltonian for heavy and light holes;
\begin{eqnarray}\label{eq:2x2Hamiltonian}
&&{\cal{H}}_{eff} = \varepsilon_{0} {\textbf{I}}-\frac{\Delta{\epsilon}}{2}\sigma_{z}\nonumber\\
&&+i\left({\lambda}'k+{\gamma}'_{1}k^3+{\gamma}'_{2}k^5+{\gamma}_{R3}k^7\right)\left(e^{-i\theta}{\sigma_{+}}-e^{i\theta}{\sigma_{-}}\right)   
\end{eqnarray}
where $\bm{k}=k_x\hat{\bm{x}}+k_y\hat{\bm{y}}$, $\varepsilon_{0}= (\epsilon_{1} + \epsilon_{2})/2$, $ \epsilon_{1(2)} = \epsilon^1_{\mathrm{HH(LH)}}+(\gamma_1+\gamma_2)\frac{\hbar^2k^2}{2m}$, $k_{\pm}=k_{x}{\pm}{i}{k_y}$, $\sigma_{\pm}=(\sigma_{x}{\pm}{i}{\sigma_y})/2$, $\Delta{\epsilon}={\epsilon}_2-{\epsilon}_1$, the third term represents the Rashba spin-orbit coefficients as ${\lambda}'=\lambda+\beta_{\chi1}{\sin{2\theta}}$, $\lambda = \frac{\sqrt{3}\hbar^2}{m}\gamma_2\int_{0}^{\infty} dzF^1_{\mathrm{HH}}(z)\frac{d}{dz}F^1_{\mathrm{LH}}(z)$, where $F^n_{\mathrm{i}}(z)$ are he envelopes given by Airy functions,  ${\gamma}'_{1}=\gamma_{R1}+\beta_{\chi2}{\sin{2\theta}}$, ${\gamma}'_{2}=\gamma_{R2}+\beta_{\chi3}{\sin{2\theta}}$ where $\theta=\arctan({k_y}/{k_x})$ is the polar angle of the wave vector $\textbf{k}$. We stress that this $2\times 2$ Hamiltonian is written in the basis $\{ 3/2, -1/2 \}$, in which one state represents heavy holes and the other light holes, using the methodology of Ref.~\cite{chow1999semiconductor}. An additional copy of this matrix exists for $\{ -3/2, 1/2 \}$. Our notation is therefore unconventional, which leads to the unusual form for the Rashba terms. The Rashba spin-orbit coefficients vary
as a function of the gate electric field (see SM III). The coefficients of Rashba SOC $\gamma_{R(n=1,2,3)}$ can be calculated as $\gamma_{Rn} =\frac{(-1)^{n}}{(2n-1)!}{{\ell}^{2n}}\left(\frac{2}{\lambda}\right)^{2n-1}$ where $\ell = -\frac{\sqrt{3}\hbar^2}{2m}\xi \gamma_2$ and $\xi = \int_{0}^{\infty} dzF^1_{\mathrm{HH}}(z)F^1_{\mathrm{LH}}(z)$. In addition, the dipolar SOC terms $\beta_{\chi{n=1,2,3}}$, arising from the non-vanishing electric-dipole matrix elements (${\cal{H}}_d$ with $\chi \ne 0$) can be written as ${\beta_{\chi{n}}} = \frac{2n}{\ell}{\gamma_{Rn}}{e{a_B}{\chi}{F_z}{\xi}}$. The term $\propto \beta_{\chi2}$ is of the same order as the Rashba spin orbit coefficients. This shows that the dipolar SOC is necessary for a quantitative theory of SOCs of hole dynamics in asymmetric GaAs QWs. This term vanishes identically when $\chi=0$. The dispersions are $
\varepsilon^{\pm}_{k}=\varepsilon_{0}\pm\sqrt{(\frac{\Delta{\epsilon}}{2})^2+\left({\lambda}'k+{\gamma}'_{1}k^3+{\gamma}'_{2}k^5+{\gamma}_{R3}k^7\right)^2}$, so that $\epsilon_{1(2)} = \varepsilon_{HH(LH)} + {(\gamma_{1}+\gamma_{2})}{\frac{\hbar^2{k^2}}{2m}}$ as shown in Fig. \ref{Fig.1}.

It has been shown that photovoltaics must be treated as a kinetic phenomenon, essentially different from the quadratic response effects caused by the interband structure \cite{sturman2020ballistic}. With this in mind, we use the quantum kinetic theory based on the density matrix \cite{PhysRevB.96.035106, PhysRevB.96.235134, PhysRevLett.124.087402}, which captures inter-band transitions induced by electric fields as well as disorder. We consider the system interacting with light, where the interaction with the time-dependent external field with monochromatic light wave is represented in length-gauge by ${\cal H}_E=-e{E_x}\cos(\omega t){{\hat {x}}}\cdot {\bf r}$. The time-dependent electric field runs along the crystal axis $(100)$. Different setups will also be considered below. We will work in the crystal momentum representation $|{s, \textbf{k}}\rangle = {e^{i{\textbf{k}\cdot{\textbf{r}}}}}|{u}^{s}_\textbf{k}\rangle$, where $|{u}^{s}_\textbf{k}\rangle$ is the lattice-periodic part of the Bloch wave function.

\begin{figure}
\includegraphics[scale=0.5]{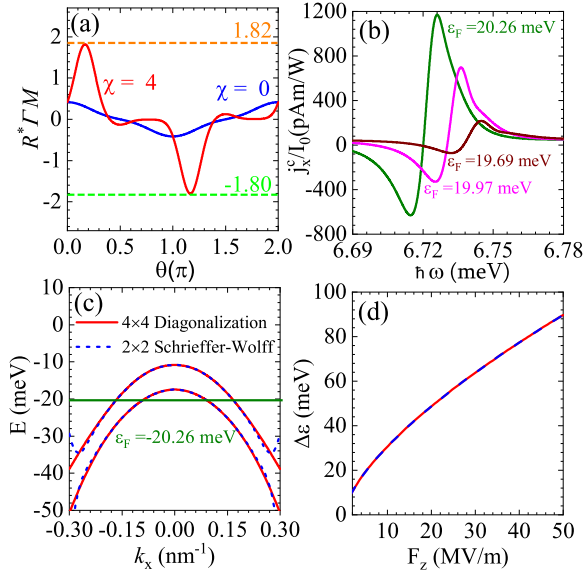}
\caption{(a) Anisotropic behavior of ${\cal R}^{*,12}_{k_x}{\Gamma}^{12}_{k_x}{\cal M}^{12}$ as a function of $\theta$ for a finite $\chi$. (b) The QEP current for the different values of the Fermi energy for the strength of scattering potential $U_0=2.5$eV m$^{-1}$ and the impurity density $n_i=2.3\times{10^{11}}$ cm$^{-2}$. The other parameters used are considered as Fig. \ref{Fig4}. (c) The HH and LH band structures in two different approaches. To have a QPE, there has to be an intersection between the Fermi energy and the LH band. (d) The variation of the LH-HH energy splitting as a function of $F_Z$. }
\label{Fig6a}
\end{figure} 

The dynamics of the density matrix (DM) are found from the quantum Liouville equation:
\begin{equation}
\frac{\partial\rho}{\partial{t}}+\frac{i}{\hbar}[{\mathcal H}_0,\rho]+J[\rho]=-\frac{i}{\hbar}[{\mathcal H}_E+{\cal{H}}_{d_x},\rho],
\end{equation} 
where the driving electric field along $\hat{x}$ gives rise to the dipolar term $H_{d_x}={\frac{1}{\sqrt{3}}}e{a_B}{\chi }{E_x}{\lbrace{J_{ y},J_{z}}\rbrace}$ and $J[\rho]$ is the scattering term. 
The single-particle DM can be decomposed into diagonal, $\rho_d$, and off-diagonal, $\rho_{od}$, parts~\cite{PhysRevLett.124.087402}. The impurity-averaged density matrix, $f=\langle\rho\rangle$, can be expanded in powers of the electric field, $f=f^{(0)}+f^{(1)}+f^{(2)}+...$ ~\cite{PhysRevLett.124.087402} and thus the quantum kinetic equation can be simplified as
\begin{eqnarray}
&&\frac{\partial f^n_d}{\partial t}+\frac{i}{\hbar}[{\mathcal H}_0,f^n_d]+J_d(f^n_d)={\cal D}[f^{(n-1)}_d]-J_{d}(f^n_{od})\\
&&\frac{\partial f^n_{od}}{\partial
t}+\frac{i}{\hbar}[{\mathcal H}_0,f^n_{od}]+J_{od}(f^n_{od})={\cal D}[f^{(n-1)}_{od}]-J_{od}(f^n_{d})\nonumber
 \end{eqnarray}
where ${\cal D}[f^{(n-1)}_{i}]=-\frac{i}{\hbar}\langle[{\mathcal H}_E+{\cal{H}}_{d_x},f^{(n-1)}_i]\rangle$ and the scattering term in the Born approximation is
\begin{equation}
J(f_{\bf k})=\frac{1}{\hbar^2}<\int_0^{\infty}dt'[{\hat U},e^{-it'{\cal H}_0/\hbar}[{\hat U},{\hat f}(t')]e^{it'{\cal H}_0\hbar}]>_{{\bf k}{\bf k}},
\end{equation}
where we assume short-ranged uncorrelated impurities with strength $U_0$ such that the average of potential over impurity configuration is $n_i |U_{{\bf k},{\bf k'}}|^2/V$ where $n_i$ is the impurity concentration and and $V$ is the crystal volume. 
We can simplify those equations by considering $J_d(f)=f/\tau_1({\bf k})$ and  $J_{od}(f)=f/\tau_2({\bf k})$ where $\tau_1$ and $\tau_2$ are the momentum relaxation times. The relaxation times account generically for impurity and phonon scattering, as well as recombination. The electric field driving term takes the form
\begin{eqnarray}\label{eq:cov}
{\cal D}[f^{(n-1)}_{od}]&&=\frac{e{\textbf{E}}}{\hbar}\cdot \frac{\partial{f^{(n-1)}_{od}}}{\partial{\textbf{k}}}-i\frac{e{\textbf{E}}}{\hbar}\cdot{\Gamma}^{ss'}_{\textbf{k}}\\
&&\times \left(f^{(n-1)}_{od}(\varepsilon^{s}(k))-f^{(n-1)}_{od}(\varepsilon^{s'}(k))\right),
\end{eqnarray}
where $\Lambda^{ss'}_{\textbf{k}}=\left(\zeta_{\textbf{k}}+(1-{\alpha}{k^2})\Omega_{\textbf{k}}\right){\hat{x}}$ with $\alpha=(\ell/\lambda)^2$ for $s\neq s'$, ${\Gamma}^{ss'}_{\textbf{k}}={\cal R}^{ss'}_{\textbf{k}}-{a_B}{\chi}{\xi}\Lambda^{ss'}_{\textbf{k}}$,  $\Omega_{\textbf{k}}=\left(\frac{{\Delta{\varepsilon}\cos{\theta}}}{2(\varepsilon_{\bm k}^{+}-\varepsilon_{0})}\right)+i{\sin{\theta}}$ and $
{\zeta}_{\textbf{k}}=\left(\frac{\beta_{k}{\Delta{\varepsilon}}k^2}{2(\varepsilon_{\bm k}^{+}-\varepsilon_{0})}\right)\left(\sqrt{(\varepsilon_{\bm k}^{+}-\varepsilon_{0})^2-(\frac{\Delta{\varepsilon}}{2})^2}-\lambda{k}\right)
$ with $\beta_{k}=\alpha/\sqrt{({\Delta{\varepsilon}}/2)^2+(\lambda{k})^2}$ [see more details on $\Lambda^{ss'}_{\textbf{k}}$ presented in SM III]. Note that the first term on the right-hand side contains the Fermi surface information, however, the second term shows the Fermi sea response and the Berry curvature information in which $\Lambda^{ss'}_{k}\neq\Lambda^{ss'}_{-k}$. The electric-dipole Hamiltonian ${\cal{H}}_{d_x}$ can be thought of as a correction to the Berry connection.  At this stage, we follow the perturbation recipe to calculate the density matrices, $f^{(1)}_{od,{\bf k}}(t)$ and $f^{(1)}_{d,{\bf k}}(t)$, to first order in the electric field. Having calculated the first order DM, the second order terms can be calculated straight away. 

The velocity tensor is given by ${\bf v}=(1/{\hbar})\left(\nabla_{\bf k} \varepsilon(k)-i[{\cal R} ,{\cal H}_0]\right)$. This includes the intraband term containing the band velocity and the interband term dependent on the Berry connection. The optical current is
$
{\bf j}^c_{s's}=\frac{-e}{\hbar}\int\frac{d{\bf k}}{4\pi^2}\nabla_{\bf k}\varepsilon^{s'}(k) f_{{\bf k}}\delta_{ss'}
+i{\cal R}^{s's}_{\bf k} [\varepsilon^{s'}(k)-\varepsilon^{s}(k)]f_{{\bf k},s's}\nonumber
$
where the first and second terms refer to intra- and inter-band contributions, respectively. The electric field needs to be incorporated into the time evolution operator leading to the scattering term. The QPE current is obtained by considering only the time-independent terms of the total current. Note that the optical current is not covered by the Fermi's golden rule since it contains only band off-diagonal elements of $r$. In other words, the Berry connection, which can also be expressed as the interband velocity. The main contribution of the optical response in our theory actually comes from $\nabla_{\bf k}$ acting on the energy in the denominator given by Eq. (S64). We stress that photovoltaic effects have a long history in non-centrosymmetric semiconductors~\cite{sturman2021photovoltaic, ivchenko2005optical}, yet all the examples studied in the past involved transitions between the valence and conduction bands, rather than between valence sub-bands, as we find here.


\begin{figure}
\centering
\includegraphics[scale=0.52]{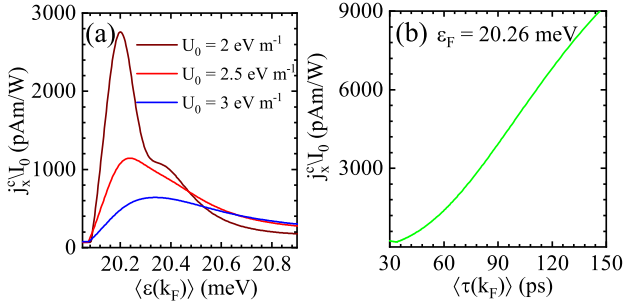}
\caption{The peak value of the photovoltaic response along the $\hat{\bm x}$ direction as a function of (a) the Fermi energy averaged over $\theta$ for different values of $U_0$ and (b) the scattering time $\langle\tau(k_{F})\rangle$ averaged over $\theta$ at $\varepsilon_F=20.26$ meV. We set $n_i=2.3\times{10^{11}}$ cm$^{-2}$ and $\tau_{1}=\tau_{2}={\tau}^a\,$, $F_z=1$ MV/m, and other parameters are given in table I in SM III.}
\label{Fig7}
\end{figure}   



\textit{Numerical Results}- Firstly, we emphasize that the anisotropy of the second-order off-diagonal DM element arises from the dipole parameter $\chi$ and hence the QPE current vanishes if the electric dipole term is neglected. We recall Fig. \ref{Fig4} where the QPE current is illustrated. In Eq. (\ref{eq:cov}), the effect of the electric dipole Hamiltonian appears as a correction to the Berry connection and leads to an angular dependence of the driving force given by $\Gamma^{ss'}_{k}$. Owing to the linear $k$ dependent of $\xi_{\bf k}$, $\Gamma^{+-}_{k}\neq \Gamma^{+-}_{-k}$ means that the driving force is no longer symmetric in ${\bf k}$ and consequently creates an imbalance in the excited population. Since ${\bf E}=E{\hat x}$ along the (100) direction,therefore, $(v_x n^{\text {eff}}(k)-v_x n^{\text {eff}}(-k))\tau$ corresponds to the displacement of excited holes. Here $v_x$ is the nearly symmetric band group velocity and $n^{\text {eff}}(k)$ is the excited hole density.
The surface energy at $\varepsilon^+_{k_F}$ oscillates under the external time-dependent electric field along $x$ ([100]) with different hole populations along $+k_x$ and $-k_x$ providing a net current that clearly depends on $\chi$ and the momentum relaxation time. Notice that this effect occurs only for hole excitation around $\varepsilon^+_{k_F}$ that describes the resonance in the QPE.
There are two optical transitions associated with two peaks given by the denominator of the QPE current terms including $(\varepsilon^{+}-\varepsilon^{-}-i\hbar\tau_{1}^{-1}-\hbar{\omega})$ and $(\varepsilon^{+}-\varepsilon^{-}-i\hbar\tau_{1}^{-1})^2-(\hbar{\omega})^2$; the direct interband transition between the LH and HH bands with the $k$ position around the intersection of the Fermi energy and the LH band, $\hbar{\omega}=\varepsilon^{+}_{k_F}-\varepsilon^{-}_{k_F}$, and another transition that consists of an intraband transition with relaxation time around the Fermi energy and then an interband transition between the LH to HH bands, $\hbar{\omega}=\varepsilon^{+}_{k_F}-\varepsilon^{-}_{k_F}-\sqrt{3/4}(\hbar\tau_1^{-1})^2/(\varepsilon^{+}_{k_F}-\varepsilon^{-}_{k_F})$. The second process is forbidden between the conduction and valence bands due to Pauli blocking in a semiconductor.

The main contribution to the QPE current originates from the second term of Eq. (S71) where ${\cal R}^{*,12}_{k_x}{\Gamma}^{12}_{k_x}{\cal M}^{12}$ is an anisotropic function with respect to the angle as shown in Fig.~\ref{Fig6a}(a) and provides a finite optical current for a finite $\chi$. Here, ${\cal M}^{ss'}({\bf k})=\left[\frac{\partial\varepsilon^{s}(k)}{\partial{k_x}}-\frac{\partial\varepsilon^{s'}(k)}{\partial{k_x}}+\frac{i\hbar}{\tau^2_2}\frac{\partial{\tau_2}}{\partial{k_x}}\right]\left[f_0(\varepsilon^{s}(k))-f_0(\varepsilon^{s'}(k))\right]$. The current is caused by the asymmetric velocity distributions in the bands as a result of kinetic processes, meaning that, if the velocity of particles for the HH and LH bands is the same, the displacement of the LH and HH Fermi surfaces is canceled out. Therefore, the QPE tends to zero. In addition, there are Fermi surface oscillations owing to the difference between the effective masses of LH and the HH bands upon optical excitation. This effect produces a resonant current peak at the interband absorption in GaAs. 

The Berry connection fulfills a key role in nonlinear optical phenomena~\cite{PhysRevLett.105.026805,PhysRevLett.115.216806}. On the other hand, the Berry connection is pertinent to the quantum geometry properties of the charge wave-function~\cite{PhysRevX.11.011001, PhysRevB.59.14915}. Quantum quantities, like the quantum metric, play a significant role in determining non-linear optical effects~\cite{morimoto2016topological,Nagaosa_NRR_NC2018,ishizuka2017local,PhysRevB.94.035117,Bhalla2021}. Our analytical calculations show that the injection current contribution~\cite{PhysRevX.11.011001,Bhalla2021}, see SM VI for the quantum metric calculations, is proportional to $|{\cal R}^{12}_{k_x}| ^2 {\cal M}^{12}({\bf k})(f^{(0)}((\varepsilon ^+)-f^{(0)}(\varepsilon ^- )))$, which represents a dominant contribution in the nonlinear optical current of the system. In addition, the higher order pole current contribution \cite{Bhalla2021} is defined by $\frac{\partial \tau }{\partial k_x}|{\cal R}^{12}_{k_x}|^ 2 (\varepsilon^+-\varepsilon^-)(f^{(0)}((\varepsilon^+)-f^{(0)}(\varepsilon ^-)))$ where $\varepsilon_i$ is on $i-$ relates to the energy band and $i=\pm$ to the HH or LH band. Our results show that both the injection contribution and the higher-order pole current make particular contributions to obtaining the nonlinear DC optical response along the $x$ and $y$ directions. Both, explained below, depend on the scattering time values, and thus the the current enhances when the relaxation time increases. However, anomalous current and double resonance current contain a negligible contribution to the DC optical response. Furthermore, we obtain the strength of the optical response depends strongly on the direction of the applied electric field.

Next, we study the effects of Fermi energy and relaxation time on the optical transitions. Increasing the Fermi energy $\varepsilon_{\text F}$ increases the Fermi surface area, resulting in a larger peak for the photovoltaic effect current, as shown in Fig. \ref{Fig6a}(b). The main peak occurs at the optical band edge. In addition, with increasing Fermi energy, there is a red shift in the photovoltaic effect current. In order to perceive the optical transitions, we concentrate on the band structure shown in Fig. \ref{Fig6a}(c) at a given Fermi energy. There is no optical response when the Fermi energy does not intersect the LH band. As $\hbar \omega$ approaches $\varepsilon^+_{k_F}-\varepsilon^-_{k_F}$ a hole can be excited from the LH sub-band to the HH sub-band. Furthermore, the optical transition point resonance varies depending on the gap between the LH and HH sub-bands. This gap can be tuned over the entire THz range (1 - 100meV) by changing the gate electric field $F_z$, Fig. \ref{Fig6a}(d), and thus the QPE can be tuned by the QW structure parameter \footnote{The model presented here, based on Airy functions, cannot go down to ultra-small gate fields, hence we have truncated the lowest energy at 1meV.}. 


We show the maximum peak value of the QPE peak along $x$ as a function of the average Fermi energy over $\theta$ in Fig. \ref{Fig7}(a) for different impurity strengths, $U_0$. First the peak decreases with increasing $U_0$. Second, since ${\cal M}^{12}({\bf k})$ is directly related to the band velocity difference between HH and LH, the current increases with increasing Fermi energy, then shows a maximum at a certain value of Fermi energy depending on the curvatures of the band structure, following which it decreases.

In Fig. \ref{Fig7}(b), the peak of the QPE current as a function of the mean relaxation time averaged over the angle $\theta$ $\langle\tau(k_F)\rangle=(1/{2\pi}) \int^{ 2\pi}_{0}{\tau(k_F, \theta)}{d{\theta}}$ for a given Fermi energy $\varepsilon_F$. 
To get the height of the peak, we identify the dominant contribution to the QPE as
\begin{eqnarray}\label{eq:approx1}
{\bf j}_{x,od}^c\propto{\frac{{\omega}\tau^2}{\hbar}}\left(\frac{\partial{f(\varepsilon^{+}(k))}}{\partial{k_x}}-\frac{\partial{f(\varepsilon^{-}(k))}}{\partial{k_x}}+\frac{i\hbar}{\tau^2}\frac{\partial{\tau}}{\partial{k_x}}\right)_{k=k_F, \theta=\pi},
 \end{eqnarray} 
 noting that only the real part is taken at the end. There is excellent agreement between the numerical calculations and Eq. (\ref{eq:approx1}). In addition, the maximum current increases quadratically with increasing $\langle\tau(k_F)\rangle$, as is clearly visible in Fig. \ref{Fig7}(b). The curve is well fitted by the formula $(\langle\tau(k_F)\rangle)^{\alpha}$, with $\alpha\sim 2$, at large and medium values of $\langle\tau(k_F)\rangle$. The slight deviation from 2 is due to computational factors.

\textit{In summary} we have considered an asymmetric structure hole GaAs QW structure and identified a strong photovoltaic response due to the quadrupolar interaction with the electric field. The size and width of the peak are determined by the momentum relaxation time, and the effect can be very strong in high-mobility systems. Since the heavy hole - light hole splitting can be tuned by a top gate field over the entire terahertz range the effect can serve as the basis for a terahertz photo-detector. Our method can be generalized to investigate the spin and orbital magnetic effect of the nonlinear optical response in GaAs QWs~\cite{wu2021anomalous,PhysRevB.83.155313,PhysRevB.90.205415}.


\acknowledgments 
D. C. is supported by the Australian Research Council Centre of Excellence in Future Low-Energy Electronics Technologies (project number CE170100039). WAC acknowledges funding from the Natural Sciences and Engineering Research Council of Canada and from the Fonds de Recherche---Nature et Technologies (Quebec).

%


\appendix

\begin{widetext}
	\section{Model Hamiltonian and photovoltaic effect}
	We intend to calculate the photovoltaic effect induced by a time-dependent in-plane electric field $\textbf{E}=E_0{\cos(\omega{t})}\hat{x}$ for a triangular GaAs quantum well. For this purpose, we consider an asymmetric quantum well-formed at a heterointerface that its confinement can be described by a triangular potential due to an electric field $\textbf{F}=F_{z}\hat{z}$,
	\begin{equation}\label{eq:triangularpotential}
		V(z) = \begin{cases} 
			\infty&\quad z\le 0,\\  
			-e F_z z&\quad z>0,\\
		\end{cases}
	\end{equation} 
	where $e=-\vert{e}\vert$ is the electron charge. For the valence band of an III-V semiconductor, where hole states are restricted to the heavy-hole (HH) and light-hole (LH) states, the full Hamiltonian matrix is given by 
	\begin{equation}\label{eq:fullH}
		{\cal{H}}={\cal{H}}_{0}+U(r)+{\cal{H}}_{E}+{\cal{H}}_{d_x},
	\end{equation}
	where $U(r)=U_0{{\sum}_i}{\delta(r-r_{i})}$ is the impurity potential and assume that the correlation function satisfies $\langle{U(r)U(r')}\rangle={n_i}{U^2_0}{\delta(r-r')}$ with $n_i$ the impurity density and $H_{E}=e\textbf{E}{\cdot}\hat{\textbf{r}}$ represents the interaction with the external electric field where $\hat{\textbf{r}}$ is the position operator and ${\cal{H}}_{d_x}$ is the effective projected electric-dipole term owing to this electric field. This term for the electric field along $\hat{x}$ is given by:
	\begin{align}
		H_{d_x}={\frac{1}{\sqrt{3}}}e{a_B}{\chi}{E_x}{\lbrace{J_{y},J_{z}}\rbrace}, 
	\end{align}
	where $e{a_{B}}\simeq2.5$D ($a_{B}$ is the Bohr radius and D is a debye), $\chi$ is a parameter that controls the strength of the electric-dipole matrix elements. Note that $J_{y}$ and $J_{z}$ are the components of the vector of spin-3/2 matrices. In a general for of an external electric field, the electric dipole is 
	$H_{d_x}={\frac{1}{\sqrt{3}}}e{a_B}{\chi}[{E_x}{\lbrace{J_{y},J_{z}}\rbrace}+{E_y}{\lbrace{J_{z},J_{x}}\rbrace}+{E_z}{\lbrace{J_{x},J_{y}}\rbrace}$.The band Hamiltonian ${\cal{H}}_{0}$ is defined as
	\begin{equation}
		{\cal{H}}_{0}={\cal{H}}_L + V(z)I_4 + {\cal{H}}_{d_z},
	\end{equation}
	where $I_4$ is the $4\times 4$ identity matrix and
	\begin{equation}\label{eq:LuttingerHam}
		{\cal{H}}_L=\frac{\hbar^2}{2m}\left[\left(\gamma_1+\frac{5}{2}\gamma_2\right)k^2 I_4-2\gamma_2(\bm{k}\cdot\mathbf{J})^2\right]
	\end{equation}
	is the Luttinger Hamiltonian within the spherical approximation, with parameters $\gamma_1=6.85$ eV and $\gamma_2=$2.10 eV.  
	Also, the effective projected electric-dipole Hamiltonian ${\cal{H}}_{d_z}$ that induced by the electrical field $\textbf{F}=F_{z}\hat{z}$ is expressed as \cite{PhysRevB.102.075310}:
	\begin{align}
		H_{d_z}={\frac{1}{\sqrt{3}}}e{a_B}{\chi}{F_z}{\lbrace{J_{x},J_{y}}\rbrace}.
	\end{align}
	In order to obtain the effective LH-HH Hamiltonian or the two dimensional hole gas, we consider $k_{x}=k_{y}=0$ in ${\cal{H}}_{0}$ so that the eigenfunction of ${\cal{H}}_0$ is the envelope function $F_{\nu}^n(z)$ solves the differential equation:
	\begin{equation}\label{eq:SE}
		\left[-\frac{\hbar^2}{2m_{\nu}}\frac{d^2}{dz^2}+ V(z)\right]F_{\nu}^n(z) = \epsilon_{\nu}^n F_{\nu}^n(z).
	\end{equation}
	Here, $m_{\nu}$ is the effective mass ($m_{\mathrm{HH}}=\frac{m}{\gamma_1-2\gamma_2}$ for heavy holes and $m_{\mathrm{LH}}=\frac{m}{\gamma_1+2\gamma_2}$ for light holes), and $\epsilon_{\nu}^n$ is the energy for subband $n$. 
	The envelopes,  $F_{\nu}^n(z)$, are given by Airy functions. 
	
	Now, we project the Hamiltonian matrix ${\cal{H}}$ onto the lowest subband ($n=1$) to obtain a $2 \times 2$ Hamiltonian matrix. For this purpose, the Hamiltonian matrix ${\cal{H}}$ can be written in terms of diagonal matrix elements of $k_z^2$, e.g.,
	\begin{equation}\label{eq:kz2}
		\left<k_z^2\right> = -\int_{0}^{\infty} dzF^1_{\mathrm{HH}}(z)\frac{d^2}{dz^2}F^1_{\mathrm{HH}}(z),
	\end{equation}
	together with the following parameters:
	\begin{equation}\label{eq:l}
		l = -\frac{\sqrt{3}\hbar^2}{2m}\xi \gamma_2
	\end{equation}
	and
	\begin{equation}\label{eq:lambda}
		\lambda = \frac{\sqrt{3}\hbar^2}{m}\gamma_2\eta_1,
	\end{equation}
	where
	\begin{equation}\label{eq:xi}
		\xi = \int_{0}^{\infty} dzF^1_{\mathrm{HH}}(z)F^1_{\mathrm{LH}}(z),
	\end{equation}
	and 
	\begin{equation}\label{eq:eta}
		\eta_i = \int_{0}^{\infty} dzF^1_{\mathrm{HH}}(z)\frac{d^i}{dz^i}F^1_{\mathrm{LH}}(z),
	\end{equation}
	where $\frac{d^i}{dz^i}$ is the $i^{\mathrm{th}}$ derivative with respect to $z$. Using second-order degenerate perturbation theory (an approximate Schrieffer-Wolff transformation \cite{PhysRevB.102.075310}), we project the Hamiltonian matrix, $\hat{H}$, onto the hole subspace [see Appendix \ref{project}].
	This gives the following effective $2 \times 2$ heavy and light holes Hamiltonian that is discussed in Appendix \ref{SWT}: 
	\begin{equation}\label{eq:2x2Hamiltonian}
		{\cal{H}}_{\text{eff}}=\left(\frac{\epsilon_{1}+\epsilon_{2}}{2}\right){\textbf{I}}-\frac{\Delta{\epsilon}}{2}\sigma_{z}+i\left({\lambda}'k+{\gamma}'_{1}k^3+{\gamma}'_{2}k^5+{\gamma}_{R3}k^7\right)\left(e^{-i\theta}{\sigma_{+}}-e^{i\theta}{\sigma_{-}}\right).    
	\end{equation}
	where $\bm{k}=k_x\hat{\bm{x}}+k_y\hat{\bm{y}}$, $ \epsilon_{1(2)} = \epsilon^1_{\mathrm{HH(LH)}}+(\gamma_1+\gamma_2)\frac{\hbar^2k^2}{2m}$, $k_{\pm}=k_{x}{\pm}{i}{k_y}$, $\sigma_{\pm}=(\sigma_{x}{\pm}{i}{\sigma_y})/2$, $\Delta{\epsilon}={\epsilon}_2-{\epsilon}_1$, $\theta=\arctan({k_y}/{k_x})$ is the polar angle of wavevector $\textbf{k}$ and the third term represents the Rashba spin-orbit coefficients as ${\lambda}'=\lambda+\beta_{\chi1}{\sin{2\theta}}$, ${\gamma}'_{1}=\gamma_{R1}+\beta_{\chi2}{\sin{2\theta}}$, ${\gamma}'_{2}=\gamma_{R2}+\beta_{\chi3}{\sin{2\theta}}$. The Rashba spin-orbit coefficients vary
	as a function of the electric field (see Fig. \ref{Fig13} in Appendix \ref{SWT}). For the range of the electric fields considered, we find that the linear Rashba spin-orbit coupling has a similar order to the cubic Rashba spin-orbit coupling. The coefficients of the Rashba spin-orbit coupling $\gamma_{R(n=1,2,3)}$ can be calculated as
	\begin{equation}\label{eq:gammaR1}
		\gamma_{Rn} =\frac{(-1)^{n}}{(2n-1)!}{{\ell}^{2n}}\left(\frac{2}{\lambda}\right)^{2n-1}.
	\end{equation} 
	In addition, the dipolar spin-orbit coupling terms $\beta_{\chi{n=1,2,3}}$ which arises from non-vanishing electric-dipole matrix elements (${\cal{H}}_d$ with $\chi \ne 0$) can be written as
	\begin{equation}\label{eq:betachi}
		{\beta_{\chi{n}}}=\frac{2n}{\ell}{\gamma_{Rn}}{e{a_B}{\chi}{F_z}{\xi}}.
	\end{equation}
	
	The term with coefficient $\beta_{\chi2}$ is of the same order as the linear Rashbar spin-orbit coefficient. This shows that the dipolar spin-orbit coupling is, therefore, necessary for a quantitative theory of the spin-orbit couplings for heavy holes in asymmetric GaAs quantum wells. Note that this term vanishes identically within the envelope-function approximation. One can calculate the dispersion relations in Eq. (\ref{eq:2x2Hamiltonian}) by $\vert{\textbf{H}-\varepsilon\textbf{I}}\vert=0$ as 
	\begin{equation}\label{eq:eigenvalues}
		\varepsilon^{\pm}_{k}=\varepsilon_{0}\pm\sqrt{(\frac{\Delta{\epsilon}}{2})^2+\left({\lambda}'k+{\gamma}'_{1}k^3+{\gamma}'_{2}k^5+{\gamma}_{R3}k^7\right)^2},
	\end{equation}
	where $\varepsilon_{0}=(\epsilon_{1}+\epsilon_{2})/2$ so that $\epsilon_{1(2)}=\varepsilon_{HH(LH)}+{(\gamma_{1}+\gamma_{2})}{\frac{\hbar^2{k^2}}{2m}}$. The eigenvectors of the system, $H_{0}\vert{u}^{s}_{\textbf{k}}\rangle={\varepsilon}^{s}_{\textbf{k}}(\textbf{k})\vert{u}^{s}_{k}\rangle$ is obtained as 
	
	\begin{eqnarray}\label{eq:eigenvectors}
		u^{s}_\textbf{k}=\frac{1}{\sqrt{2f(k, \theta)}}
		\begin{pmatrix}
			s\sqrt{f(k, \theta)-s\frac{\Delta{\epsilon}}{2}} \\
			-i\sqrt{f(k, \theta)+s\frac{\Delta{\epsilon}}{2}}e^{i{\theta}}
		\end{pmatrix}, s=\pm{1},
	\end{eqnarray}
	where $f(k,\theta)$ equals to
	\begin{equation}\label{eq:f(k)}
		f(k, \theta)=\sqrt{(\frac{\Delta{\epsilon}}{2})^2+\left({\lambda}'k+{\gamma}'_{1}k^3+{\gamma}'_{2}k^5+{\gamma}_{R3}k^7\right)^2}. 
	\end{equation}
	In order to calculate the Berry connection, we do need to calculate the ${{\nabla}_{\textbf{k}}}{u^{s'}_{\textbf{k}}}$. To do so, we make use of the polar coordinate as
	\begin{eqnarray}
		\nabla_{\textbf{k}}{u^{s'}_{\textbf{k}}}={\partial_{k}}{\vert{u}^{s'}_{k}\rangle}{\hat{k}}+{\frac{1}{k}}{\partial_{\theta}}{\vert{{u}^{s'}_{k}}\rangle}{\hat{\theta}}
		={\frac{\Delta{\epsilon}{\sqrt{2f}}{\partial_{k}f}}{8f^2}}\begin{pmatrix}
			\frac{1}{\sqrt{f-s'\frac{\Delta{\epsilon}}{2}}} \\
			i\frac{s'{e^{i\theta}}}{\sqrt{f+s'\frac{\Delta{\epsilon}}{2}}}
		\end{pmatrix}{\hat{k}}+\frac{\Delta{\epsilon}\sqrt{2f}{\partial_{\theta}f}}{8{k}f^2}\begin{pmatrix}
			\frac{1}{\sqrt{f-s'\frac{\Delta{\epsilon}}{2}}}  \\
			\left(\frac{is'}{\sqrt{f+s'\frac{\Delta{\epsilon}}{2}}}+\frac{4f}{{\partial_{\theta}f}\Delta{\epsilon}}{\sqrt{f+s'\frac{\Delta{\epsilon}}{2}}}\right){e^{i\theta}}\end{pmatrix}
		{\hat{\theta}},
	\end{eqnarray}
	where
	\begin{eqnarray}
		\partial_{k}{f}(k, \theta)=\frac{1}{f(k, \theta)}\left({\lambda}'+3{\gamma}'_{1}k^2+5{\gamma}'_{2}k^4+7{\gamma}_{R3}k^6\right)\left({\lambda}'k+{\gamma}'_{1}k^3+{\gamma}'_{2}k^5+{\gamma}_{R3}k^7\right),\nonumber\\ 
		\partial_{\theta}{f}(k, \theta)=2{\cos{2\theta}}\frac{\left({\beta_{\chi1}}k+{\beta_{\chi2}}k^3+{\beta_{\chi3}}k^5\right)}{f(k, \theta)}\left({\lambda}'k+{\gamma}'_{1}k^3+{\gamma}'_{2}k^5+{\gamma}_{R3}k^7\right).
	\end{eqnarray}
	The Berry connection part for different band indices (i.e, $s=+$,  $s'=-$) is
	\begin{eqnarray}
		{\cal R}^{+-}_{\textbf{k}}={i}\langle{u}^{-}_k|\nabla_{k}|{u}^{+}_k\rangle
		=i\frac{\Delta{\epsilon}\left({\lambda}'+3{\gamma}'_{1}k^2+5{\gamma}'_{2}k^4+7{\gamma}_{R3}k^6\right)}{4f^2}{\hat{k}}
		-{\frac{\left({\lambda}'+{\gamma}'_{1}k^2+{\gamma}'_{2}k^4+{\gamma}_{R3}k^6\right)}{2f}}{\hat{\theta}}\nonumber\\ 
		+i\frac{\Delta{\epsilon}{\cos{2\theta}}\left(\beta_{\chi1}+\beta_{\chi2}k^2+\beta_{\chi3}k^4\right)}{2f^2}{\hat{\theta}}.  
	\end{eqnarray}
	Notice that the condition ${\cal R}^{-+}_{\textbf{k}}={\cal R}^{*,+-}_{\textbf{k}}$ is satisfied. Using the relationship between unit vectors in Cartesian and polar coordinates $\hat{k}=\hat{x}\cos{\theta}+\hat{y}\sin{\theta}$ and $\hat{\theta}=-\hat{x}\sin{\theta}+\hat{y}\cos{\theta}$, one can write Berry connections along the $\hat{x}$ and $\hat{y}$ directions as ${\cal R}^{+-}_{k_x}=R_{1}{\cos{\theta}}+R_{2}{\sin{\theta}}$ and ${\cal R}^{+-}_{k_y}=R_{1}{\sin{\theta}}-R_{2}{\cos{\theta}}$,  respectively. Here $R_{1}$ and $R_{2}$ as follows   
	\begin{eqnarray}
		{R_1}=i\frac{\Delta{\epsilon}\left({\lambda}'+3{\gamma}'_{1}k^2+5{\gamma}'_{2}k^4+7{\gamma}_{R3}k^6\right)}{4f^2},\nonumber\\
		{R_2}={\frac{\left({\lambda}'+{\gamma}'_{1}k^2+{\gamma}'_{2}k^4+{\gamma}_{R3}k^6\right)}{2f}}-i\frac{\Delta{\epsilon}{\cos{2\theta}}\left(\beta_{\chi1}+\beta_{\chi2}k^2+\beta_{\chi3}k^4\right)}{2f^2}.
	\end{eqnarray}
	We also consider a short-range (onsite) disorder of the symmetric form $U(\textbf{r}) =U_0\sum_i\delta({\bf r}-{\bf r}_i)$ so that the matrix elements of $U^{ss'}_{\textbf{k}\textbf{k}'}$ is defined as
	\begin{eqnarray}
		U^{ss'}_{\textbf{k}\textbf{k}'}&&=\langle{s, \textbf{k}}|U(\textbf{r})|{s',\textbf{k}' }\rangle ={U_0}\langle{u}^{s}_\textbf{k}|{u}^{s'}_{\textbf{k}'}\rangle{\int}{d\textbf{r}}{\sum_{\textbf{R}}{e^{-i(\textbf{k}-\textbf{k}')\cdot{\textbf{r}}}}\delta({\bf r}-\textbf{R})}={U_0}\langle{u}^{s}_\textbf{k}|{u}^{s'}_{\textbf{k}'}\rangle. 
	\end{eqnarray}
	Note that $|{s,\textbf{k} }\rangle={e^{i{\textbf{k}\cdot{\textbf{r}}}}}{{u}^{s}_\textbf{k}}$ is a Bloch wave function. Here,$\langle{u}^{s}_\textbf{k}|{u}^{s'}_{\textbf{k}'}\rangle$ as follows
	\begin{eqnarray}
		\langle{u}^{s}_\textbf{k}|{u}^{s'}_{\textbf{k}'}\rangle=\frac{1}{2\sqrt{f(k, \theta)f(k', {\theta}')}}\left({ss'}{\sqrt{f(k, {\theta})-s{\frac{\Delta{\epsilon}}{2}}}}{\sqrt{f(k', {\theta}')-s'{\frac{\Delta{\epsilon}}{2}}}}+{\sqrt{f(k, {\theta})+s{\frac{\Delta{\epsilon}}{2}}}}{\sqrt{f(k', {\theta}')+s'{\frac{\Delta{\epsilon}}{2}}}}{e^{i\gamma}}\right).
		\nonumber
	\end{eqnarray}
	where $\gamma=\theta'-\theta$. The second-order term in the scattering potential can be written as follows:
	\begin{eqnarray}
		\langle U^{sm}_{\bf kk'}U^{ms'}_{\bf k'k}\rangle=\frac{U^2_0}{V}{\langle{u}^{s}_\textbf{k}|{u}^{m}_{\textbf{k}'}\rangle}{\langle{u}^{m}_\textbf{k}|{u}^{s'}_{\textbf{k}'}\rangle}\sum_{\textbf{R}_1, \textbf{R}_2}{e^{-i(\textbf{k}-\textbf{k}')\cdot({\textbf{R}_1-\textbf{R}_2})}}.
	\end{eqnarray}
	Using translation symmetry, this disorder average depends only on $\textbf{R}={\textbf{R}_1-\textbf{R}_2}$. It follows that 
	\begin{eqnarray}
		\langle U^{sm}_{\bf kk'}U^{ms'}_{\bf k'k}\rangle=\frac{U^2_0}{V}{\langle{u}^{s}_\textbf{k}|{u}^{m}_{\textbf{k}'}\rangle}{\langle{u}^{m}_\textbf{k}|{u}^{s'}_{\textbf{k}'}\rangle}{N}{\sum_{\textbf{R}}{e^{-i(\textbf{k}-\textbf{k}')\cdot{\textbf{R}}}}}={n_i}{U^2_0}{\langle{u}^{s}_\textbf{k}|{u}^{m}_{\textbf{k}'}\rangle}{\langle{u}^{m}_\textbf{k}|{u}^{s'}_{\textbf{k}'}\rangle}
	\end{eqnarray}
	It is assumed that the correlation function satisfies $\langle{U(\textbf{r})U(\textbf{r}')}\rangle={n_i}{U^2_0}{\delta{(\textbf{r}-\textbf{r}')}}$ with $n_i$ the impurity density, so the relaxation times $\tau^{s}_{\pm}$ for these impurities via the Fermi's golden rule is calculated as [see more details in Appendix \ref{RT}]  
	\begin{eqnarray}
		\frac{1}{\tau^{s}_{\pm}}=\frac{2{\pi}}{\hbar}{\sum_{m}}\int {\frac{k'{dk'}{d{\theta'}}}{(2\pi)^2}} \langle U^{sm}_{\bf kk'}U^{ms'}_{\bf k'k}\rangle\delta(\varepsilon^{\pm}(k')-\varepsilon^{\pm}(k))={\frac{n_i{U^2_0}}{2\pi\hbar}}\int{k'{dk'}{d{\theta}'}}{\delta(\varepsilon^{\pm}(k')-\varepsilon^{\pm}(k))}\nonumber\\
		\approx{\frac{1}{2\pi\tau_{0}}}{\int}{d{\theta'}}\left(1\pm{\frac{B}{2A{k}^2}}\mp{\frac{\partial_{k}f}{2A{k}}}-{\frac{3B{\partial_{k}f}}{2A^2{k}^3}}+{\frac{3B^2}{4A^2{k}^4}}+\frac{1}{2}(\frac{\partial_{k}f}{A{k}})^2+\frac{B\partial^2_{k}f}{2A^2{k}^2}\right),
	\end{eqnarray}
	where $\tau_{0}={\hslash^3}(\gamma_{1}+\gamma_{2}){/}{({n_i}m{U^2_0})}$, $A={\hslash^{2}}(\gamma_{1}+\gamma_{2})/{2m}$ and $B=f(k, \theta')-f(k, \theta)$. The average relaxation time for a symmetric scattering potential can be written as
	\begin{eqnarray}
		\frac{1}{{\tau}^{s}(k, \theta)}=\frac{1}{2}\sum_{s}\frac{1}{{\tau_{s}}}={\frac{1}{\tau_{0}}}\left[1+{\frac{1}{2\pi}}{\int}d{\theta}'\left(-{\frac{3B{\partial_{k}f}(k, \theta')}{2A^2{k}^3}}+{\frac{3B^2}{4A^2{k}^4}}+\frac{1}{2}(\frac{\partial_{k}f(k,\theta')}{A{k}})^2+\frac{B\partial^2_{k}f(k, \theta')}{2A^2{k}^2}\right)\right].
	\end{eqnarray}
	We have so far studied the contribution of the symmetric scattering potential. There will be an asymmetric scattering potential when the excitation and scattering are associated with distinct defects. In this case, we consider an asymmetric potential as $U(r, \theta) ={U_0}{\sum_i}({r-r_i})\cos{\theta}$, so the matrix elements of $U^{ss'}_{\textbf{k}\textbf{k}'}$ is calculated as
	\begin{eqnarray}
		U^{ss'}_{\textbf{k}\textbf{k}'}&&=\langle{s,\textbf{k}}|U(\textbf{r})|{s',\textbf{k}'}\rangle=\langle{s,\textbf{k}}|{U_0}{\sum_i}{(r-r_i)}\cos{\varphi}''|{s',\textbf{k}'}\rangle={U_0}\langle{u}^{s}_{\textbf{k}}|{u}^{s'}_{\textbf{k}'}\rangle{\sum_{R}}{\int}{d{\textbf{r}}}{e^{i{(\textbf{k}'-\textbf{k})\cdot{\textbf{r}}}}}{(r-R){\cos{\varphi}''}}.
	\end{eqnarray}
	\begin{figure}[ht]
		\centering
		\includegraphics[scale=0.35]{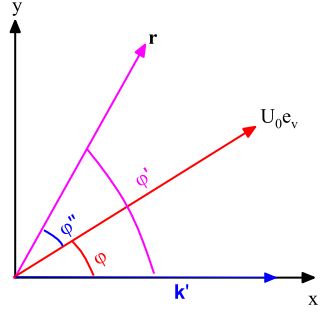}
		\caption{Schematic illustration from different angles in the scattering problem. The only physical angle here is that between the ejected electron momentum $\textbf{k}'$ and the direction of scattering potential $U_0\hat{e_v}$ because we expect the electron to be ejected maximally along $U_0\hat{e_v}$. This suggests rearranging the axes to have the electron momentum along the $x$-axis and $\theta$ being the angle between the electron momentum $\textbf{k}'$ and the direction of scattering potential. Also, we will consider $\textbf{k}\parallel{\textbf{r}}$.}
		\label{Fig.2}
	\end{figure} 
	Here, $|{s,\textbf{k} }\rangle={e^{i{\textbf{k}\cdot{\textbf{r}}}}}{{u}^{s}_\textbf{k}}$ is a Bloch wave function. The above equation can be written as follows:
	\begin{eqnarray}
		U^{ss'}_{\textbf{k}\textbf{k}'}&&={U_0}\langle{u}^{s}_{\textbf{k}}|{u}^{s'}_{\textbf{k}'}\rangle{\sum_{R}}{\int}{d{\textbf{r}}}{e^{i{(\textbf{k}'-\textbf{k})\cdot{(\textbf{r}+\textbf{R})}}}}{r{\cos{\varphi}''}}={U_0}\langle{u}^{s}_{\textbf{k}}|{u}^{s'}_{\textbf{k}'}\rangle{\sum_{R}}{e^{i{(\textbf{k}'-\textbf{k})\cdot{\textbf{R}}}}}{\int}{d{\textbf{r}}}{e^{i{(\textbf{k}'-\textbf{k})\cdot{\textbf{r}}}}}r{\cos{\varphi}''}.
	\end{eqnarray}
	According to Fig. \ref{Fig.2}, $\cos{\varphi}''$ can be written as 
	\begin{equation}  
		\cos{\varphi}''=\cos({\varphi}'-{\varphi})=\cos{\varphi}'\cos{\varphi}+\sin{\varphi}'\sin{\varphi}.
	\end{equation}
	The final equation can be hence written as
	\begin{equation}
		U^{ss'}_{\textbf{k}\textbf{k}'}={U_0}\langle{u}^{s}_{\textbf{k}}|{u}^{s'}_{\textbf{k}'}\rangle{\int}{r^2{dr}}{\int^{2\pi}_0}{e^{{-ikr}}}\left(\cos{\varphi}'\cos{\varphi}+\sin{\varphi}'\sin{\varphi}\right){e^{i{k'{{r\cos{\varphi}'}}}}}{d{\varphi}'}={U_0}\langle{u}^{s}_{\textbf{k}}|{u}^{s'}_{\textbf{k}'}\rangle{\cos{\varphi}}{2i\pi}{\int}r^2{e^{-ikr}{J_1({k}'r)}{dr}}.
	\end{equation}
	The contribution of the second sentence is zero because ${\int^{2\pi}_0}\sin{\varphi}'{e^{i{k'{{r\cos{\varphi}'}}}}}{d{\varphi}'}=0$ and considering elastic scattering (i.e, $k=k'$) and substituting $\vert{k'-k}\vert\sim{k}$, the integral in the last equation is solved for $k\ll$ as
	\begin{equation}
		{\int}{r^2{e^{-ikr}}{J_1({k}'r)}{dr}}\approx\frac{-2}{k^3}.  
	\end{equation}
	Finally, the matrix elements of the scattering potential are
	\begin{equation}
		U^{sm}_{\textbf{k}\textbf{k}'}U^{ms'}_{\textbf{k}'\textbf{k}}=\left(\frac{4{\pi}U_0}{k^3}\right)^2 \langle{u}^{s}_{\textbf{k}}|{u}^{m}_{\textbf{k}'}\rangle{\langle{u}^{m}_{\textbf{k}'}|{u}^{s'}_{\textbf{k}}\rangle}\cos^{2}{\varphi}.
	\end{equation}  
	For the second order of scattering potential, one can prove that $\langle{U^{sm}_{\textbf{k}\textbf{k}'}U^{ms'}_{\textbf{k}'\textbf{k}}}\rangle=n_i{U^{sm}_{\textbf{k}\textbf{k}'}U^{ms'}_{\textbf{k}'\textbf{k}}}$. Thus, the average of matrix elements of the scattering potential is
	\begin{eqnarray}
		\langle{U^{sm}_{\textbf{k}\textbf{k}'}U^{ms'}_{\textbf{k}'\textbf{k}}}\rangle={n_i}\frac{4{\pi}^2U_0^2}{k^6}\langle{u}^{s}_{\textbf{k}}|{u}^{m}_{\textbf{k}'}\rangle{\langle{u}^{m}_{\textbf{k}'}|{u}^{s'}_{\textbf{k}}\rangle},
	\end{eqnarray}  
	where we employ $\langle{\cos^{2}{\varphi}}\rangle=1/2$. The time relaxation via Fermi's golden rule for an asymmetric scattering potential is given as
	\begin{eqnarray}
		\frac{1}{\tau^{a}_{\pm}}=\frac{2{\pi}}{\hbar}{\sum_{m}}\int {\frac{k'{dk'}{d{\theta'}}}{(2\pi)^2}}\langle{U^{sm}_{\bf kk'}U^{ms'}_{\bf k'k}}\rangle\delta(\varepsilon^{\pm}(k')-\varepsilon^{\pm}(k))\nonumber\\
		\approx\frac{{\pi}{n_i}U_0^2}{A{\hbar}k^6}{\int}{d{\theta'}}\left(1\pm{\frac{B}{2A{k}^2}}\mp{\frac{\partial_{k}f}{2A{k}}}-{\frac{3B{\partial_{k}f}}{2A^2{k}^3}}+{\frac{3B^2}{4A^2{k}^4}}+\frac{1}{2}(\frac{\partial_{k}f}{A{k}})^2+\frac{B\partial^2_{k}f}{2A^2{k}^2}\right)
	\end{eqnarray}
	where $A={\hslash^{2}}(\gamma_{1}+\gamma_{2})/{2m}$ and $B=f(k, \theta')-f(k, \theta)$. Ignoring the small term $B^2$, the average relaxation time is calculated as
	\begin{eqnarray}
		\frac{1}{{\tau}^{a}(k, \theta)}=\frac{1}{2}\sum_{s}\frac{1}{{\tau^{a}_{s}}}\approx\frac{{\pi}{n_i}U_0^2}{A{\hbar}k^6}{\int^{2\pi}_0}{d{\theta'}}\left(1-{\frac{3B{\partial_{k}f}}{2A^2{k}^3}}+\frac{1}{2}(\frac{\partial_{k}f}{A{k}})^2+\frac{B\partial^2_{k}f}{2A^2{k}^2}\right).
	\end{eqnarray}
	If we ignore the small terms including $\chi^3$ and $\chi^4$, the final result of the above integral can be expressed as
	\begin{eqnarray}
		&&\frac{1}{{\tau}^{a}(k, \theta)}\approx{\frac{1}{\tau_{0}(k)}}\Bigg(1-\frac{3}{2A^2k^3}\left[f_1{\partial_{k}f_1}+\frac{1}{2}f_2{\partial_{k}f_2}-{f(k, \theta)\sqrt{(\frac{\Delta{\varepsilon}}{2})^2+f^2_1}}\left(\frac{\partial_{k}f_1}{f_1}-\frac{\partial_{k}f_2}{f_2}{(\frac{\Delta{\varepsilon}}{2f_1})^2}\right)\right]\nonumber\\
		&+&\frac{1}{2(Ak)^2}\Bigg[(\partial_{k}f_1)^2+{f_1{\partial^2_{k}f_1}}+\frac{1}{2}(f_2{\partial^2_{k}f_2}+({\partial_{k}f_2})^2)-{f(k, \theta)\sqrt{(\frac{\Delta{\varepsilon}}{2})^2+f^2_1}}\Bigg(\frac{\partial^2_{k}f_1}{f_1}+\frac{\partial^2_{k}f_2}{f_2}+2\frac{\partial_{k}f_1}{f_1}\frac{\partial_{k}f_2}{f_2}\nonumber\\
		&+&\frac{1}{9}\left((\frac{\partial_{k}f_1}{f_1}+\frac{\partial_{k}f_2}{f_2})^2+2\frac{\partial_{k}f_2}{f_2}\frac{\partial_{k}f_1}{f_1}\right)-\frac{((\partial_{k}f_2)^2+f_2\partial_{k}f_2)[(\frac{\Delta{\varepsilon}}{2})^2+f^2_1]}{(f_1f_2)^2}-\frac{2}{3}\frac{({\partial_{k}f_1})^2}{[(\frac{\Delta{\varepsilon}}{2})^2+f^2_1]}(1+\frac{f_1\partial_{k}f_2}{f_2\partial_{k}f_1})\Bigg)\Bigg]\Bigg)  
	\end{eqnarray}
	where $f_1={\lambda}k+{\gamma_{R1}}k^3+{\gamma_{R2}}k^5+{\gamma_{R3}}k^7$,$f_2={\beta_{\chi1}}k+{\beta_{\chi2}}k^3+{\beta_{\chi3}}k^5$ and $\tau_{0}(k)={A{\hbar}k^6}/{2{\pi}^2{n_i}U_0^2}$. Finally, the total average relaxation time equals to $1/{\tau}=1/{{\tau}^{a}}+1/{{\tau}^{s}}$ and then we will consider $\tau_{1}=\tau_{2}={\tau}$. Figure \ref{Fig3} shows the relaxation time due to the asymmetric scattering potential as a function of $\theta$ at different values of $k$. As we see below, there is an asymmetry about $\theta$ and its magnitude changes with respect to $k$.
	\begin{figure}[ht]
		\centering
		\includegraphics[scale=0.7]{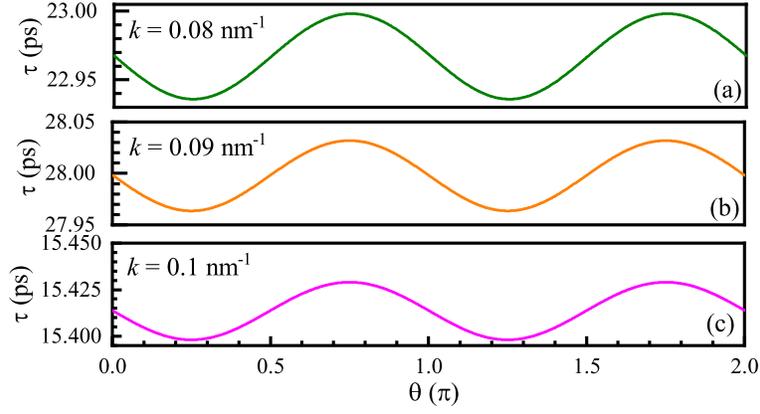}
		\caption{The relaxation time due to the asymmetric scattering potential with the strength of $U_0=4$ eV m$^{-1}$ at (a) $k=0.08$ nm$^{-1}$, (b) $k=0.09$ nm$^{-1}$ and (c) $k=0.1$ nm$^{-1}$.}
		\label{Fig3}
	\end{figure} 
	\subsection{Photovoltaic current}
	
	Based on the DM equation, the density matrix is given by $\rho=| \psi \rangle \langle \psi |$. The dynamic of the density matrix obeys quantum Liouville equation:
	\begin{equation}
		\frac{\partial\rho}{\partial{t}}+\frac{i}{\hbar}[{\mathcal H}_0,\rho]+J[\rho]=-\frac{i}{\hbar}[{\mathcal H}_E+{\cal{H}}_{d_x},\rho],
	\end{equation} 
	where $J[\rho]$ is the scattering term which takes the form with in the Born approximation and we assume the correlation function $\langle U({\bf r})U({\bf r}')\rangle=n_i U_0^2\delta({\bf r}-{\bf r}')$ with $n_i$ the impurity density. Hence, the general form of $J[x]$ is
	\begin{eqnarray}\label{eq:J}
		J[\langle\rho\rangle]^{ss'}_{\bf k}&&=\frac{\pi}{\hbar} \sum_{mm',{\bf k'}} \{ \langle U^{sm}_{\bf kk'}U^{mm'}_{\bf k'k}\rangle\langle\rho\rangle^{m's'}_{\bf k}\delta(\varepsilon^m(k')-\varepsilon^{m'}(k))+\langle U^{mm'}_{\bf kk'}U^{m's'}_{\bf k'k}\rangle\langle\rho\rangle^{sm}_{\bf k}\delta(\varepsilon^m(k)-\varepsilon^{m'}(k'))\nonumber\\
		&&-\langle U^{sm}_{\bf kk'}U^{m's'}_{\bf k'k}\rangle\langle\rho\rangle^{mm'}_{\bf k'}\delta(\varepsilon^{m'}(k')-\varepsilon^{s'}(k))
		-\langle U^{sm}_{\bf kk'}U^{m's'}_{\bf k'k}\rangle\langle\rho\rangle^{mm'}_{\bf k'}\delta(\varepsilon^{s}(k)-\varepsilon^{m}(k'))\}.
	\end{eqnarray}
	In order to obtain that expression we make use of $\pi \delta(x-a)=\int_0^{\infty} dk e^{ik(x-a)}$. We define the average of the density matrix over impurity configurations as $f^t=\langle \rho \rangle$ where $f^t_{\bf k}=f_{{k}}$ in which $f_{{k}}$ can be decomposed into two parts namely a diagonal matrix in band index, $f_d$ and an off-diagonal term, $f_{od}$ which we subsequently have $f_{k}=f_{d,{\bf k}} \oplus f_{od,{\bf k}}$. Therefore,
	\begin{eqnarray}
		f^t_{\bf k}=\langle | \psi \rangle \langle \psi | \rangle=
		\begin{pmatrix}
			f_d^{c,+}& f_{od}^{+}&0&0 \\
			f_{od}^{*+}&f_d^{v,+}&0&0\\
			0& 0 &f_d^{c,-}&f_{od}^{-}\\
			0&0&f_{od}^{*-}&f_d^{v,-}
		\end{pmatrix},
	\end{eqnarray} 
	where $\pm$ refers to pseudospin and $HH(LH)$ denotes the heavy (light) band. According to Eq. (\ref{eq:J}),  we can decompose the scattering term into diagonal and off-diagonal parts as
	\begin{eqnarray}
		[J_{d}(f_{d})]^{ss}_{\bf k}&&=\frac{2\pi}{\hbar}\sum_{m,{\bf k'}}{\langle U^{sm}_{\bf kk'}U^{ms}_{\bf k'k}\rangle}(f^{ss}_{d,{\bf k}}-f^{mm}_{d,{\bf k'}})\delta(\varepsilon^s(k)-\varepsilon^{m}(k')),
	\end{eqnarray}
	\begin{eqnarray}
		[J_{od}(f_{d})]^{sm'}_{\bf k}&&=\frac{\pi}{\hbar}\sum_{m,{\bf k'}}{\langle U^{sm}_{\bf kk'}U^{mm'}_{\bf k'k}\rangle}\left((f^{ss}_{d,{\bf k}}-f^{mm}_{d,{\bf k'}})\delta(\varepsilon^s(k)-\varepsilon^{m}(k'))+(f^{m'{m'}}_{d,{\bf k}}-f^{mm}_{d,{\bf k'}})\delta(\varepsilon^{m'}(k)-\varepsilon^{m}(k'))\right). 
	\end{eqnarray}
	To get the corresponding expressions for the band off-diagonal part of the density matrix, $f_{od}$, we employ the following expression which was shown recently in detail in the second-order nonlinear Hall effect \cite{winkler2003spin}
	\begin{eqnarray}
		[J_{d}(f_{od})]^{ss}_{\bf k}=\frac{\pi}{\hbar} \sum_{mm',{\bf k'}} \{\langle U^{sm}_{\bf kk'}U^{mm'}_{\bf k'k}\rangle{f}^{m's}_{{od},{\bf k}}\delta(\varepsilon^{m}(k')-\varepsilon^{m'}(k))+\langle U^{mm'}_{\bf kk'}U^{m's}_{\bf k'k}\rangle{f}^{sm}_{{od},{\bf k}}\delta(\varepsilon^{m}(k)-\varepsilon^{m'}(k'))\nonumber\\
		-\langle U^{sm}_{\bf kk'}U^{m's}_{\bf k'k}\rangle{f}^{m{m'}}_{{od},{\bf k}}\delta(\varepsilon^{m'}(k')-\varepsilon^{s}(k))
		-\langle U^{sm}_{\bf kk'}U^{m's}_{\bf k'k}\rangle{f}^{m{m'}}_{{od},{\bf k}}\delta(\varepsilon^{s}(k)-\varepsilon^{m}(k'))\}
	\end{eqnarray}
	The density matrix can be expanded in the powers of the electric field and thus the quantum kinetic equation can be simplified as
	\begin{eqnarray}
		&&\frac{\partial f^n_d}{\partial t}+\frac{i}{\hbar}[{\mathcal H}_0,f^n_d]+J(f^{t n})=-\frac{i}{\hbar}[{\mathcal H}_E+{\cal{H}}_{d_x},f^{(n-1)}_d],\\
		&&\frac{\partial f^n_{od}}{\partial
			t}+\frac{i}{\hbar}[{\mathcal H}_0,f^n_{od}]+J(f^{t n})=-\frac{i}{\hbar}[{\mathcal H}_E+{\cal{H}}_{d_x},f^{(n-1)}_{od}].\nonumber\\
	\end{eqnarray}
	
	Therefore, we will get

	\begin{eqnarray}
		&&\frac{\partial f^n_d}{\partial t}+\frac{i}{\hbar}[{\mathcal H}_0,f^n_d]+J_d(f^n_d)=-\frac{i}{\hbar}[{\mathcal H}_E+{\cal{H}}_{d_x},f^{(n-1)}_d]-J_{d}(f^n_{od}),\\
		&&\frac{\partial  f^n_{od}}{\partial t}+\frac{i}{\hbar}[{\mathcal H}_0,f^n_{od}]+J_{od}(f^n_{od})=-\frac{i}{\hbar}[{\mathcal H}_E+{\cal{H}}_{d_x},f^{(n-1)}_{od}]-J_{od}(f^n_{d}).\nonumber\\
	\end{eqnarray}
	We can simplify those equations by considering $J_d(f)=f/\tau_1$ and  $J_{od}(f)=f/\tau_2$, where $\tau_1$ and $\tau_2$ are relaxation times. The final equations for the density matrix become
	\begin{eqnarray}
		&&\frac{\partial f^n_d}{\partial t}+\frac{i}{\hbar}[{\mathcal H}_0,f^n_d]+\frac{f^n_d}{\tau_1}=-\frac{i}{\hbar}[{\mathcal H}_E+{\cal{H}}_{d_x},f^{(n-1)}_d]-J_{d}(f^n_{od}),\\
		&&\frac{\partial f^n_{od}}{\partial t}+\frac{i}{\hbar}[{\mathcal H}_0,f^n_{od}]+\frac{f^n_{od}}{\tau_2}=-\frac{i}{\hbar}[{\mathcal H}_E+{\cal{H}}_{d_x},f^{(n-1)}_{od}]-J_{od}(f^n_{d}),\nonumber\\
	\end{eqnarray}
	Meanwhile we also use the covariant derivative where
	\begin{eqnarray}\label{eq:D}
		-\frac{i}{\hbar} \langle [H_{E}+{\cal{H}}_{d_x}, f^{(n-1)}]\rangle=\frac{e{\textbf{E}}}{\hbar}\cdot\left[\frac{\partial{f^{(n-1)}}}{\partial{\textbf{k}}}-i({\cal R}^{ss'}_{\textbf{k}}-{a_B}{\chi}{\xi}\Lambda^{ss'}_{\textbf{k}})\left(f^{(n-1)}(\varepsilon^{s}(k))-f^{(n-1)}(\varepsilon^{s'}(k))\right)\right].
	\end{eqnarray}
	Here, $\Lambda^{ss'}_{\textbf{k}}=\left(\zeta_{\bm k}+(1-\alpha{k^2})\Omega_{\bm k}\right){\hat{x}}$ with $\alpha=(\ell/\lambda)^2$ [more details about matrix $\Lambda^{ss'}_{\textbf{k}}$ is presented in Appendix \ref{SWT}]. It is found that the effect of electric-dipole Hamiltonian ${\cal{H}}_{d_x}$ appears as a correction in Berry connection according to Eq. (\ref{eq:D}). Then, we defined ${\Gamma}^{ss'}_{\textbf{k}}={\cal R}^{ss'}_{\textbf{k}}-{a_B}{\chi}{\xi}\Lambda^{ss'}_{\textbf{k}}$ for simplicity. Notice that we will consider the external electric field of a form $\textbf{E}(t)={E_0}{\cos{\omega{t}}}\,{\hat{x}}$. In this stage we follow the perturbation recipe to calculate the first order density matrices, $f^{(1)}_{od,{\bf k}}(t)$ and $f^{(1)}_{d,{\bf k}}(t)$. They are given by 
	\begin{equation}
		f^{(1),{ss}}_{d,{\bf k}}(t)=\int_{-\infty}^t ~dt' e^{-\frac{t-t'}{\tau_1}} \left[\frac{e{E(t')}}{\hbar}\frac{\partial{f_0(\varepsilon^{s}(k))}}{\partial{k_x}}-J_{d}[(f^{(1)}_{od,{\bf k}}(t'))]^{ss}\right]
	\end{equation}
	and
	\begin{equation}
		f^{(1),{ss'}}_{od,{\bf k}}(t)=\int_{-\infty}^t ~dt' e^{-\frac{t-t'}{\tau_2}} e^{\frac{-{i}\varepsilon^{s}(k)(t-t')}{\hbar}}\left[-\frac{e{E(t')}}{\hbar}\left(i{\Gamma}^{ss'}_{k_{x}} [f_0(\varepsilon^{s}(k))-f_0(\varepsilon^{s'}(k))\right)-[J_{od}(f^{(1)}_{d,{\bf k}}(t'))]^{ss'}\right] e^{\frac{i\varepsilon^{s'}(k)(t-t')}{\hbar}},
	\end{equation}
	where $f_0(\varepsilon^s(k))$ is an equilibrium Fermi-Dirac distribution function. 
	Therefore, other terms can be given by
	\begin{eqnarray}
		&&f^{(2),{ss}}_{d,{\bf k}}(t)=\int_{-\infty}^t ~dt' e^{-\frac{t-t'}{\tau_1}}\left[\frac{e{E(t')}}{\hbar}\left(\frac{f^{(1),ss}_{d,{\bf k}}(t')}{\partial_{k_{x}}}-i[{\Gamma}^{ss}_{k_{x}}, \langle\rho\rangle^{(1)}]^{ss}\right)-[J_{d}(f^{(2)}_{od,{\bf k}}(t'))]^{ss}\right],\nonumber\\
		&&f^{(2)}_{od,{\bf k}}(t)=\int_{-\infty}^t ~dt' e^{-\frac{t-t'}{\tau_2}} e^{-i\frac{\varepsilon^{s}(k)(t-t')}{\hbar}}\left[\frac{e{E(t')}}{\hbar}\left(\frac{f^{(1)}_{od,{\bf k}}(t')}{\partial{k_{x}}}-i[{\Gamma}^{ss'}_{k_{x}}, \langle\rho\rangle^{(1)}]^{ss'}\right)-[J_{od}(f^{(2)}_{d,{\bf k}}(t'))]^{ss'}\right]e^{i\frac{\varepsilon^{s'}(k)(t-t')}{\hbar}}.\nonumber\\
	\end{eqnarray}
	Here, we would like to simplify those expressions by making use of the following relation:
	\begin{equation}
		[{\Gamma}^{ss'}_{k_{x}}, \langle\rho\rangle^{(1)}]^{ss'}=f^{(1),ss'}_{od,{\bf k}}(t)({\Gamma}^{ss}_{k_{x}}-{\Gamma}^{s's'}_{k_{x}})+{\Gamma}^{ss'}_{k_{x}}(f^{(1),s's'}_{d,{\bf k}}(t)-f^{(1),ss}_{d,{\bf k}}(t)),
	\end{equation}
	we consider the nonzero terms of the diagonal $f$ when $s=s'$ and the off-diagonal $f$ when $s \neq s'$.
	The optical current will be obtained through the following expression: 
	\begin{equation}
		{\bf j}^{c}_{ss'}=-{e}\int \frac{d{\bf k}}{4\pi^2} \langle{s,\textbf{k}}|{\bf v}{\bf f}^t_{{\bf k}}|{s',\textbf{k}}\rangle,
	\end{equation}
	where the velocity tensor ${\bf v}=(1/{\hbar})D {\cal H}_{0}/D{\bf k}$ and $|{s,\textbf{k}}\rangle={e^{i{\textbf{k}\cdot{\textbf{r}}}}}{{u}^{s}_\textbf{k}}$ is a Bloch wave function. By expanding the covariant derivative, the velocity is given by ${\bf v}=(1/{\hbar})\left(\nabla_{\bf k} \varepsilon(k)-i[{\cal R} ,{\cal H}_0]\right)$
	and eventually the total current is
	\begin{equation}
		{\bf j}^c_{s's}=-\frac{e}{\hbar}\int\frac{d{\bf k}}{4\pi^2}\left(\nabla_{\bf k}\varepsilon^{s'}(k) f^t_{{\bf k}}\delta_{ss'}+i{\cal R}^{s's}_{\bf k} [\varepsilon^{s'}(k)-\varepsilon^{s}(k)]f^t_{{\bf k},s's}\right).
	\end{equation}
	It can be easily proven from ${\bf j}^{c}_{s's}={\bf j}^{*,c}_{ss'}$ that ${\cal R}^{s's}_{\bf k}={\cal R}^{*,ss'}_{\bf k}$. The x-and $y$-components of the off-diagonal part of the photovoltaic current will be
	\begin{eqnarray}
		{\bf j}_{x,od}^c=+i\frac{e}{\hbar}\int \frac{d{\bf k}}{4\pi^2} {{\cal R}^{*{ss'}}_{k_x}} [\varepsilon^{s}(k)-\varepsilon^{s'}(k)]f^{(2),{ss'}}_{od,{\bf k}},\nonumber\\
		{\bf j}_{y,od}^c=+i\frac{e}{\hbar}\int \frac{d{\bf k}}{4\pi^2} {{\cal R}^{*{ss'}}_{k_y}} [\varepsilon^{s}(k)-\varepsilon^{s'}(k)]f^{(2),{ss'}}_{od,{\bf k}}.
	\end{eqnarray}
	Also, the $x$- and $y$-components of the diagonal part of the current can be written as
	\begin{eqnarray}
		{\bf j}_{x,d}^c=-\frac{e}{\hbar}\int \frac{d{\bf k}}{4\pi^2}\frac{\partial{\varepsilon^{s}(k)}}{\partial{k_x}}f^{(2),ss}_{d,\bf k},\nonumber\\
		{\bf j}_{y,d}^c=-\frac{e}{\hbar}\int \frac{d{\bf k}}{4\pi^2}\frac{\partial{\varepsilon^{s}(k)}}{\partial{k_y}}f^{(2),ss}_{d,\bf k}.
	\end{eqnarray}
	
	\subsection{The time-dependent electric field}
	Now, we consider the external time-dependent field as ${\bf E}={E_0}{\cos(\omega{t})}{\hat x}$. By substituting that in the density matrix formula, they yield as
	\begin{eqnarray}
		f^{(1),ss}_{d,{\bf k}}(t)=\int_{-\infty}^t ~dt' e^{-\frac{t-t'}{\tau_1}}\left[{\frac{eE_0}{\hslash}}{(\frac{e^{i{\omega}{t'}}+e^{-i{\omega}{t'}}}{2})}{\frac{\partial{f_0(\varepsilon^{s}(k))}}{\partial{k_x}}}-[J_{d}(f^{(1)}_{od,{\bf k}}(t'))]^{ss}\right],   
	\end{eqnarray}
	\begin{eqnarray}
		f^{(1),ss}_{d,{\bf k}}(t)=\sum_{l=\pm}{\frac{eE_{0}}{2\hbar}}\frac{e^{il{\omega}{t}}}{\frac{1}{\tau_1}+il{\omega}}
		{\frac{\partial{f_0(\varepsilon^{s}(k))}}{\partial{k_x}}}-\int_{-\infty}^t ~dt' e^{-\frac{t-t'}{\tau_1}}[J_{d}(f^{(1)}_{od,{\bf k}}(t'))]^{ss}
	\end{eqnarray}
	\begin{eqnarray}
		f^{(1),ss'}_{\text{od},{\bm k}}(t)=-\sum_{l=\pm}\frac{e{E_{0}}}{2}\frac{e^{il \omega t}{\Gamma}^{ss'}_{k_x}[f_0(\varepsilon^{s}(k))-f_0(\varepsilon^{s'}(k)]}{l\hbar\omega+\varepsilon^{s}(k)-\varepsilon^{s'}(k)-i\frac{\hbar}{\tau_{2}}}-\int_{-\infty}^t ~dt' e^{-\frac{t-t'}{\tau_2}} e^{-i\frac{\varepsilon^{s}(k)(t-t')}{\hbar}}[J_{od}(f^{(1)}_{d,{\bf k}}(t'))]^{ss'}e^{i\frac{\varepsilon^{s'}(k)(t-t')}{\hbar}}.
	\end{eqnarray}
	Now, we calculate $\partial{f^{(1)}_{d,{\bf k}}(t)}/{\partial{k_x}}$
	\begin{eqnarray}
		\frac{\partial{f^{(1),{ss}}_{d,{\bf k}}(t)}}{\partial{k_x}}=\sum_{l=\pm}{\frac{eE}{2\hbar}}{\frac{\partial{\tau_1}}{\partial{k_x}}}\frac{e^{il{\omega}{t}}}{(1+il{\omega}{\tau_{1}})^2}{\frac{\partial{f_0(\varepsilon^{s}(k))}}{\partial{k_x}}}+\sum_{l=\pm}{\frac{eE}{2\hbar}}\frac{e^{il{\omega}{t}}}{\frac{1}{\tau_1}+il{\omega}}{\frac{\partial^{2}{f_0(\varepsilon^{s}(k))}}{\partial{k_x^{2}}}}\nonumber\\
		-\int_{-\infty}^t ~dt' e^{-\frac{t-t'}{\tau_1}}\frac{\partial{[J_{d}(f^{(1)}_{od,{\bf k}}(t'))]^{ss}}}{\partial{k_x}}-\int_{-\infty}^t ~dt' \frac{(t-t')}{\tau^{2}_{1}}{(\frac{\partial{\tau_1}}{\partial{k_x}})}e^{-\frac{t-t'}{\tau_1}}{{[J_{d}(f^{(1)}_{od,{\bf k}}(t'))]^{ss}}}.
	\end{eqnarray}
	In the same manner we have
	\begin{eqnarray}
		\frac{\partial{f^{(1),ss'}_{od,{\bf k}}(t)}}{\partial{k_x}}=-\sum_{l=\pm}{\frac{e{E_0}}{2}}\frac{e^{il \omega t}{\Gamma}^{ss'}_{k_x}{\cal P}^{ss'}({\bf k})}{l\hbar\omega+\varepsilon^{s}(k)-\varepsilon^{s'}(k)-i\frac{\hbar}{\tau_{2}}}+\sum_{l=\pm}\frac{e{E_0}}{2}\frac{e^{il \omega t}{\Gamma}^{ss'}_{k_x}{\cal M}^{ss'}({\bf k})}{\left({l\hbar\omega+\varepsilon^{s}(k)-\varepsilon^{s'}(k)-i\frac{\hbar}{\tau_{2}}}\right)^2}\nonumber
	\end{eqnarray}
	\begin{eqnarray}
		-\sum_{l=\pm}\frac{e{E_0}}{2}\frac{e^{il \omega t}{\partial_{k_x}{\Gamma}^{ss'}_{k_x}}[f_0(\varepsilon^{s}(k))-f_0(\varepsilon^{s'}(k)]}{l\hbar\omega+\varepsilon^{s}(k)-\varepsilon^{s'}(k)-i\frac{\hbar}{\tau_{2}}}\nonumber
	\end{eqnarray}
	\begin{eqnarray}
		+i\int_{-\infty}^t ~dt' (\frac{t-t'}{\hbar}) e^{-\frac{t-t'}{\tau_2}}\frac{\partial{\varepsilon^{s}(k)}}{\partial{k_x}}e^{-i\frac{\varepsilon^{s}(k)(t-t')}{\hbar}}{[J_{od}(f^{(1)}_{d,{\bf k}}(t'))]^{ss'}}e^{i\frac{\varepsilon^{s'}(k)(t-t')}{\hbar}}\nonumber
	\end{eqnarray}
	\begin{eqnarray}
		-\int_{-\infty}^t ~dt' e^{-\frac{t-t'}{\tau_2}}e^{-i\frac{\varepsilon^{s}(k)(t-t')}{\hbar}}{\frac{\partial[J_{od}(f^{(1)}_{d,{\bf k}}(t'))]^{ss'}}{\partial{k_x}}}e^{i\frac{\varepsilon^{s'}(k)(t-t')}{\hbar}}\nonumber
	\end{eqnarray}
	\begin{eqnarray}
		-i\int_{-\infty}^t ~dt' (\frac{t-t'}{\hbar}) e^{-\frac{t-t'}{\tau_2}}e^{-i\frac{\varepsilon^{s}(k)(t-t')}{\hbar}}[J_{od}(f^{(1)}_{d,{\bf k}}(t'))]^{ss'}{\frac{\partial\varepsilon^{s'}(k)}{\partial{k_x}}}e^{i\frac{\varepsilon^{s'}(k)(t-t')}{\hbar}},\nonumber
	\end{eqnarray} 
	\begin{eqnarray}
		-\int_{-\infty}^t ~dt' (\frac{t-t'}{\tau^{2}_{2}})(\frac{\partial{\tau_{2}}}{\partial{k_x}})e^{-\frac{t-t'}{\tau_2}}e^{-i\frac{\varepsilon^{s}(k)(t-t')}{\hbar}}[J_{od}(f^{(1)}_{d,{\bf k}}(t'))]^{ss'}e^{i\frac{\varepsilon^{s'}(k)(t-t')}{\hbar}},
	\end{eqnarray}
	where we define 
	\begin{equation}
		{\cal P}^{ss'}({\bf k})=\left[\frac{\partial{f_0(\varepsilon^{s}(k))}}{\partial{k_x}}-\frac{\partial{f_0(\varepsilon^{s'}(k))}}{\partial{k_x}}\right],\nonumber\\
	\end{equation}
	\begin{equation}
		{\cal M}^{ss'}({\bf k})=\left[\frac{\partial\varepsilon^{s}(k)}{\partial{k_x}}-\frac{\partial\varepsilon^{s'}(k)}{\partial{k_x}}+\frac{i\hbar}{\tau^2_2}\frac{\partial{\tau_2}}{\partial{k_x}}\right]\left[f_0(\varepsilon^{s}(k))-f_0(\varepsilon^{s'}(k))\right].
	\end{equation}
	In addition, the second-order density matrix is given by
	\begin{equation}
		f^{(2),ss}_{d,{\bf k}}(t)=\int_{-\infty}^t ~dt' e^{-\frac{t-t'}{\tau_1}}\left[\frac{eE(t')}{\hbar}{\frac{\partial f^{(1)}_{d,{\bf k}}(t')}{\partial{k_x}}}-[J_{d}(f^{(2)}_{od,{\bf k}}(t'))]^{ss}\right]
	\end{equation}
	\begin{eqnarray}
		&&f^{(2),ss}_{d,{\bm k}}(t) = (\frac{eE_{0}}{2\hbar})^2\bigg\{\sum_{l=\pm}\frac{e^{2il\omega t}}{(il\omega + \tau_{1}^{-1})(2il\omega +\tau_{1}^{-1})}+ \frac{2}{\omega^{2} + \tau_{1}^{-2}} \bigg\}{\frac{\partial^{2}{f_0(\varepsilon^{s}(k))}}{\partial{k_x^{2}}}}\nonumber\\
		&+&(\frac{eE_{0}}{2\hbar})^2\bigg\{\sum_{l=\pm}\frac{e^{2il\omega t}}{(il\omega\tau_{1}+1)^2(2il\omega +\tau_{1}^{-1})}+ \frac{\tau_{1}}{(1+il\omega\tau_{1})^2} \bigg\}{\frac{\partial{f_0(\varepsilon^{s}(k))}}{\partial{k_x}}}\frac{\partial{\tau_{1}}}{\partial{k_x}}\nonumber\\
		&-&\int^t_{-\infty} dt' e^{-\frac{t-t'}{\tau_{1}}}\bigg\{\frac{e{E_0}}{\hbar}\bigg(\frac{e^{i\omega t'}+e^{-i\omega t'}}{2}\bigg)\bigg[\int_{-\infty}^{t'} dt'' e^{-\frac{t'-t''}{\tau}}\frac{\partial{[J_\text{d}(f_{\text{od},{\bm k}}^{(1)}(t''))]^{ss}}}{\partial{k_x}}\bigg]\bigg\}\nonumber\\ 
		&-&\int_{-\infty}^{t} dt' e^{-\frac{t-t'}{\tau_{1}}}[J_{\text{d}}(f_{\text{od},{\bm k}}^{(2)}(t'))]^{ss}-\int^t_{-\infty} dt' e^{-\frac{t-t'}{\tau_{1}}}\bigg\{\frac{e{E_0}}{\hbar}{\frac{\partial{\tau_1}}{\partial{k_x}}}\bigg(\frac{e^{i\omega t'}+e^{-i\omega t'}}{2}\bigg)\bigg[\int_{-\infty}^{t'} dt''\frac{(t'-t'')}{\tau^{2}_{1}} e^{-\frac{t'-t''}{\tau}}{[J_\text{d}(f_{\text{od},{\bm k}}^{(1)}(t''))]^{ss}}\bigg]\bigg\}.
	\end{eqnarray}
	We can also write the off-diagonal term as
	$$
	f^{(2),ss'}_{od,{\bf k}}(t)=\int_{-\infty}^t ~dt' e^{-\frac{t-t'}{\tau_2}} e^{-i\frac{\varepsilon^{s}(k)(t-t')}{\hbar}}\left[\frac{e{E(t')}}{\hbar}\{\frac{\partial{f^{(1),ss'}_{od,{\bf k}}(t')}}{\partial{k_x}}-i[{\Gamma}_{k_x}^{ss'}, \langle\rho\rangle^{(1)}]^{ss'}\}-[J_{od}(f^{(2)}_{d,{\bf k}}(t'))]^{ss'}\right]e^{i\frac{\varepsilon^{s'}(k)(t-t')}{\hbar}}
	$$
	\begin{eqnarray}
		&&f^{(2)}_{od,{\bf k}}(t)=(\frac{eE_0}{2})^2\bigg\{\sum_{l=\pm}\frac{e^{i 2l\omega t}}{i({\varepsilon^{s}-\varepsilon^{s'}+2l{\hbar}\omega})+\frac{\hbar}{\tau_{2}}}+\frac{2}{\frac{\hbar}{\tau_{2}}+i({\varepsilon^{s}-\varepsilon^{s'}})}\bigg\}\nonumber\\
		&&\bigg\{-\frac{{\Gamma}^{ss'}_{k_x}{\cal P}^{ss'}({\bf k})}{l\hbar\omega+\varepsilon^{s}(k)-\varepsilon^{s'}(k)-i\frac{\hbar}{\tau_{2}}}+\frac{{\Gamma}^{ss'}_{k_x}{\cal M}^{ss'}({\bf k})}{\left(l\hbar\omega+\varepsilon^{s}(k)-\varepsilon^{s'}(k)-i\frac{\hbar}{\tau_{2}}\right)^2}-\frac{{\partial_{k_x}{\Gamma}^{ss'}_{k_x}}[f_0(\varepsilon^{s}(k))-f_0(\varepsilon^{s'}(k)]}{l\hbar\omega+\varepsilon^{s}(k)-\varepsilon^{s'}(k)-i\frac{\hbar}{\tau_{2}}}+\frac{i{\cal P}^{ss'}{\Gamma}^{ss'}_{k_x}}{\frac{\hbar}{\tau_{1}}+il\hbar\omega}\bigg\}\nonumber\\
		&+&\int_{-\infty}^t ~dt' e^{-\frac{t-t'}{\tau_2}} e^{-i\frac{\varepsilon^{s}(k)(t-t')}{\hbar}}\left[\frac{e{E(t')}}{\hbar}\left(F_1(t')+F_2(t')+F_3(t')+F_4(t')\right)\right]e^{i\frac{\varepsilon^{s'}(k)(t-t')}{\hbar}}\nonumber\\
		&+i&\int_{-\infty}^t ~dt' e^{-\frac{t-t'}{\tau_2}}e^{-i\frac{\varepsilon^{s}(k)(t-t')}{\hbar}}\left[\frac{e{E(t')}}{\hbar}\left(F^{s{s}}_5(t')-F^{s'{s'}}_5(t')\right)\right]{{\Gamma}^{ss'}_{k_x}}e^{i\frac{\varepsilon^{s'}(k)(t-t')}{\hbar}}\nonumber\\
		&-&\int_{-\infty}^t ~dt' e^{-\frac{t-t'}{\tau_2}} e^{-i\frac{\varepsilon^{s}(k)(t-t')}{\hbar}}[J_{od}(f^{(2)}_{d,{\bf k}}(t'))]^{ss'} e^{i\frac{\varepsilon^{s'}(k)(t-t')}{\hbar}},\nonumber\\
	\end{eqnarray}
	where
	\begin{eqnarray}
		&&F_1(t')=+i\int_{-\infty}^{t'}~dt'' (\frac{t'-t''}{\hbar}) e^{-\frac{t'-t''}{\tau_2}}{\frac{\partial{\varepsilon^{s}(k)}}{\partial{k_x}}}e^{-i\frac{\varepsilon^{s}(k)(t'-t'')}{\hbar}}[J_{od}(f^{(1)}_{d,{\bf k}}(t''))]^{ss'}e^{i\frac{\varepsilon^{s'}(k)(t'-t'')}{\hbar}},\nonumber\\
		&&F_2(t')=-\int_{-\infty}^{t'} ~dt'' e^{-\frac{t'-t''}{\tau_2}}e^{-i\frac{\varepsilon^{s}(k)(t'-t'')}{\hbar}}\frac{\partial{[J_{od}(f^{(1)}_{d,{\bf k}}(t''))]^{ss'}}}{\partial{k_x}}e^{i\frac{\varepsilon^{s'}(k)(t'-t'')}{\hbar}},\nonumber\\
		&&F_3(t')=-i\int_{-\infty}^{t'} ~dt'' (\frac{t'-t''}{\hbar}) e^{-\frac{t'-t''}{\tau_2}}e^{-i\frac{\varepsilon^{s}(k)(t'-t'')}{\hbar}}[J_{od}(f^{(1)}_{d,{\bf k}}(t''))]^{ss'}\frac{\partial{\varepsilon^{s'}(k)}}{\partial{k_x}}e^{i\frac{\varepsilon^{s'}(k)(t'-t'')}{\hbar}},\nonumber\\
		&&F_4(t')=-\int_{-\infty}^{t'} ~dt'' (\frac{t'-t''}{\tau^{2}_{2}})(\frac{\partial{\tau_{2}}}{\partial{k_x}})e^{-\frac{t'-t''}{\tau_2}}e^{-i\frac{\varepsilon^{s}(k)(t'-t'')}{\hbar}}[J_{od}(f^{(1)}_{d,{\bf k}}(t''))]^{ss'}e^{i\frac{\varepsilon^{s'}(k)(t'-t'')}{\hbar}},\nonumber\\
		&&F^{s{s}}_5(t')=\int_{-\infty}^{t'} ~dt'' e^{-\frac{t'-t''}{\tau_1}}[J_{d}(f^{(1)}_{od,{\bf k}}(t''))]^{ss}.\nonumber\\
	\end{eqnarray}
	
	The $x$-component of the off-diagonal part of the current will be
	\begin{equation}
		{\bf j}_{x,od}^c=+i\frac{e}{\hbar}\int \frac{d{\bf k}}{4\pi^2}{\cal R}^{*,ss'}_{k_x} [\varepsilon^{s}(k)-\varepsilon^{s'}(k)]f^{(2),ss'}_{od,{\bf k}}.
	\end{equation}
	In addition, we are mainly keen on time independent term contribution which leads to the DC current.
	\begin{eqnarray}
		{\bf j}_{x,od}^c=i\frac{e}{\hbar}\int \frac{kd{k}d{\theta}}{4\pi^2}(\frac{eE_0}{2})^2\bigg\{\frac{2}{\frac{\hbar}{\tau_{2}}+i({\varepsilon^{s}-\varepsilon^{s'}})}\bigg\}{\sum_{l=\pm}}\bigg\{-\frac{{\cal R}^{*,ss'}_{k_x}{\Gamma}^{ss'}_{k_x}{\cal P}^{ss'}({\varepsilon^{s}-\varepsilon^{s'}})}{l\hbar\omega+\varepsilon^{s}-\varepsilon^{s'}-i\frac{\hbar}{\tau_{2}}}\nonumber\\
		+\frac{{\cal R}^{*,ss'}_{k_x}{\Gamma}^{ss'}_{k_x}{\cal M}^{ss'}({\varepsilon^{s}-\varepsilon^{s'}})}{(l\hbar\omega+\varepsilon^{s}-\varepsilon^{s'}-i\frac{\hbar}{\tau_{2}})^2}-\frac{{\cal R}^{*,ss'}_{k_x}{\partial_{k_x}{\Gamma}^{ss'}_{k_x}}(f_0(\varepsilon^{s}(k))-f_0(\varepsilon^{s'}(k))({\varepsilon^{s}-\varepsilon^{s'}})}{l\hbar\omega+\varepsilon^{s}-\varepsilon^{s'}+i\frac{\hbar}{\tau_{2}}}+\frac{i{\cal P}^{ss'}{\cal R}^{*,ss'}_{k_x}{\Gamma}^{ss'}_{k_x}(\varepsilon^{s}-\varepsilon^{s'})}{\frac{\hbar}{\tau_{1}}+il\hbar\omega}\bigg\}\nonumber\\
		-e\int \frac{kd{k}d{\theta}}{4\pi^2}{\cal R}^{*,ss'}_{k_x}\frac{[J_{od}(f^{(2)}_{d,{\bf k}})]^{ss'}(\varepsilon^{s}-\varepsilon^{s'})}{(\varepsilon^{s}-\varepsilon^{s'})-i\frac{\hbar}{\tau_2}} 
	\end{eqnarray}
	The above equations is rewritten as
	\begin{eqnarray}\label{eq:jx}
		&&{\bf j}_{x,od}^c=-\frac{e}{\hbar}\int \frac{kd{k}d{\theta}}{4\pi^2}({eE_0})^2\bigg\{\frac{\left({\cal R}^{*,ss'}_{k_x}{\Gamma}^{ss'}_{k_x}{\cal P}^{ss'}+{{\cal R}^{*}}_{k_x}{\partial_{k_x}{\Gamma}_{k_x}}(f_0(\varepsilon^{s})-f_0(\varepsilon^{s'}))\right)(\varepsilon^{s}-\varepsilon^{s'})}{\left(\varepsilon^{s}-\varepsilon^{s'}-i\frac{\hbar}{\tau_{2}}\right)^2-(\hbar{\omega})^2}\bigg\}\nonumber\\
		&+&\frac{e}{\hbar}\int \frac{kd{k}d{\theta}}{4\pi^2}({eE_0})^2\bigg\{\frac{{\cal R}^{*,ss'}_{k_x}{\Gamma}^{ss'}_{k_x}{\cal M}^{ss'}\left((\varepsilon^{s}-\varepsilon^{s'}-i\frac{\hbar}{\tau_{2}})^2+(\hbar{\omega})^2\right) (\varepsilon^{s}-\varepsilon^{s'})}{\left(\left(\varepsilon^{s}-\varepsilon^{s'}-i\frac{\hbar}{\tau_{2}}\right)^2-(\hbar{\omega})^2\right)^{2}\left(\varepsilon^{s}-\varepsilon^{s'}-i\frac{\hbar}{\tau_{2}}\right)}\bigg\}\nonumber\\
		&-&\frac{e}{\hbar}\int \frac{kd{k}d{\theta}}{4\pi^2}({eE_0})^2\bigg\{\frac{{\cal P}^{ss'}{\cal R}^{*,ss'}_{k_x}{\Gamma}^{ss'}_{k_x}(\frac{\hbar}{\tau_{1}})(\varepsilon^{s}-\varepsilon^{s'})}{\left((\frac{\hbar}{\tau_{1}})^2+(\hbar{\omega})^2\right)\left(\frac{\hbar}{\tau_{2}}+i(\varepsilon^{s}-\varepsilon^{s'})\right)}\bigg\}-e\int \frac{kd{k}d{\theta}}{4\pi^2}{\cal R}^{*,ss'}_{k_x}\frac{[J_{od}(f^{(2)}_{d,{\bf k}})]^{ss'}(\varepsilon^{s}-\varepsilon^{s'})}{(\varepsilon^{s}-\varepsilon^{s'})-i\frac{\hbar}{\tau_2}}.  
	\end{eqnarray}
	Now, we employ the following approximation [see Appendix \ref{RT}]: 
	\begin{equation}
		\frac{\partial{f_{0}(\varepsilon^{s})}}{\partial{\varepsilon^{s}}}\approx-\delta(\varepsilon^{s}-\varepsilon_{F})\approx-\left(1+s{\frac{m(f(k)-f(k_F))}{\hbar^{2}k(\gamma_{1}+\gamma_{2})}}\frac{\partial}{\partial{k}}\right)\delta\left(\varepsilon_{0}(k)-\varepsilon_{0}(k_F)\right),
	\end{equation}
	where $\varepsilon_{0}(k)=\frac{\varepsilon_{HH}+\varepsilon_{LH}}{2}+\frac{\hslash^{2}}{2m}(\gamma_{1}+\gamma_{2}){k^2}$.  ${\cal P}^{ss'}$ can be approximated in the above equation as
	\begin{equation}
		{\cal P}^{ss'}\approx-{\frac{m}{\hbar^{2}k(\gamma_{1}+\gamma_{2})}}\left[\frac{\partial{\varepsilon^{s}(k)}}{\partial{k_x}}-\frac{\partial{\varepsilon^{s'}(k)}}{\partial{k_x}}\right]\delta(k-k_{F})+\left({\frac{m}{\hbar^{2}k(\gamma_{1}+\gamma_{2})}}\right)^2{\partial_{k}f}\left[s\frac{\partial{\varepsilon^{s}(k)}}{\partial{k_x}}-s'\frac{\partial{\varepsilon^{s'}(k)}}{\partial{k_x}}\right]\delta(k-k_F).
	\end{equation}
	Here, ${\partial{\varepsilon^{s}(k)}}/{\partial{k_x}}$ in the polar coordinate system is given by
	\begin{eqnarray}
		\frac{\partial{\varepsilon^{s}(k)}}{\partial{k_x}}={\frac{\hslash^2{k}(\gamma_{1}+\gamma_{2})}{m}}\cos{\theta}+s\left({\partial_{k}f}\cos{\theta}-\frac{\sin{\theta}{\partial_{\theta}f}}{k}\right).
	\end{eqnarray}
	
	In addition, the Fermi occupation number difference factor is  $f_0(\varepsilon^{s}(k)){\approx} f_0(\varepsilon_0)+sf{\partial_{\varepsilon_0}}f_0(\varepsilon_0)$ can be written approximately as
	\begin{equation}
		f_0(\varepsilon^{s}(k))-f_0(\varepsilon^{s'}(k))\approx{\frac{m(s'-s)}{\hbar^{2}k(\gamma_{1}+\gamma_{2})}}f(k)\delta(k-k_{F}).
	\end{equation}
	
	In order to calculate the $y$-component of the off-diagonal part of the current, it is enough to replace ${\cal R}_{k_x}$ with ${\cal R}_{k_y}$ in Eq. (\ref{eq:jx}). 
	
	Here, the $x$-component of the diagonal part of the photovoltaic current can be written as
	\begin{equation}
		{\bf j}_{x,d}^c=-\frac{e}{\hbar}\int \frac{d{\bf k}}{4\pi^2}\frac{\partial{\varepsilon(k)}}{\partial{k_x}}f^{(2)}_{d,\bf k}.
	\end{equation}
	According to our approximations in this report, it will be as
	\begin{equation}
		{\bf j}_{x,d}^c=-\frac{e^3{E^{2}_0}}{2\hbar}\int \frac{k{dk}{d\theta}}{(4\pi^2)\left((\hbar{\omega})^2+(\frac{\hbar}{\tau_{1}})^2\right)}{\frac{\partial{\varepsilon(k)}}{\partial{k_x}}}{\frac{\partial^{2}{f_0(\varepsilon^{s}(k))}}{\partial{k_x^{2}}}}+\frac{e^3{E^{2}_0}}{2\hbar}\int\frac{k{dk}{d\theta}\left((\hbar{\omega})^2-(\frac{\hbar}{\tau_{1}})^2\right)}{(4\pi^2{\tau_1})\left((\hbar{\omega})^2+(\frac{\hbar}{\tau_{1}})^2\right)^2}{\frac{\partial{\varepsilon(k)}}{\partial{k_x}}}{\frac{\partial{f_0(\varepsilon^{s}(k))}}{\partial{k_x}}}\frac{\partial{\tau_{1}}}{\partial{k_x}}.
	\end{equation}
	The above equation is simplified as
	\begin{eqnarray}
		&&{\bf j}_{x,d}^c=\frac{e^3{E_0^2}}{2\hbar}\int \frac{k{dk}{d\theta}}{(4\pi^2)\left((\hbar{\omega})^2+(\frac{\hbar}{\tau_{1}})^2\right)}\frac{\partial{\varepsilon^{s}(k)}}{\partial{k_x}}\left[\frac{\partial^2{\varepsilon^{s}(k)}}{\partial{k_x}^2}-\frac{\left((\hbar{\omega})^2-(\frac{\hbar}{\tau_{1}})^2\right)}{{\tau_1}\left((\hbar{\omega})^2+(\frac{\hbar}{\tau_{1}})^2\right)}\frac{\partial{\varepsilon^{s}(k)}}{\partial{k_x}}\frac{\partial{\tau_{1}}}{\partial{k_x}}\right]\delta(\varepsilon^{s}(k)-\varepsilon(k_F))\nonumber\\
		&+&\frac{e^3{E_0^2}}{2\hbar}\int\frac{k{dk}{d\theta}}{(4\pi^2)\left((\hbar{\omega})^2+(\frac{\hbar}{\tau_{1}})^2\right)} \left(\frac{\partial{\varepsilon^{s}(k)}}{\partial{k_x}}\right)^2{\frac{\partial{\delta(\varepsilon^{s}(k)-\varepsilon(k_F))}}{\partial{k_{x}}}},
	\end{eqnarray}
	where 
	\begin{eqnarray}
		\frac{\partial^2{\varepsilon^{s}(k)}}{\partial{k_x}^2}=\frac{\hslash^{2}}{m}(\gamma_{1}+\gamma_{2})+s\left({\partial^2_{k}f}\cos^{2}{\theta}+\frac{1}{k}(\partial_{k}f+\frac{\partial^{2}_{\theta}f}{k})\sin^{2}{\theta}+\frac{\sin{2\theta}}{2k}(\frac{2\partial_{\theta}f}{k}-\partial_{\theta}\partial_{k}f-\partial_{k}\partial_{\theta}f)\right),
	\end{eqnarray}
	\begin{eqnarray}
		{\frac{\partial{\delta(\varepsilon^{s}(k)-\varepsilon(k_F))}}{\partial{k_{x}}}}=\cos{\theta}\left(\frac{\partial}{\partial{k}}+\frac{sm(k{\partial_{k}f}-f(k)-f(k_F))}{\hslash^{2}k^2(\gamma_{1}+\gamma_{2})}{\frac{\partial}{\partial{k}}}+\frac{sm(f(k)-f(k_F))}{\hslash^{2}k(\gamma_{1}+\gamma_{2})}{\frac{\partial^2}{\partial{k^2}}}\right)\delta^{0}-\frac{ms{\partial_{\theta}f}\sin{\theta}}{(\hslash{k})^2(\gamma_{1}+\gamma_{2})}{\frac{\partial{\delta}^{0}}{\partial{k}}}, 
	\end{eqnarray}
	
	where $\delta^{0}=\delta(\varepsilon_{0}(k)-\varepsilon_{0}(k_F))$. Then, we employ $f(x){\delta^{(n)}(x)}={(-1)^{n}}{f^{(n)}(x)}{\delta(x)}$ for obtaining the $x$-component of the diagonal part of the photovoltaic current. Notice that for calculating the $y$-component of the diagonal part of current, one can only replace ${\partial{\varepsilon^{s}(k)}}/{\partial{k_y}}$ with ${\partial{\varepsilon^{s}(k)}}/{\partial{k_x}}$ in the first term of above equation and $\left({\partial{\varepsilon^{s}(k)}}/{\partial{k_x}}\right)^2$ wtih $({\partial{\varepsilon^{s}(k)}}/{\partial{k_y}})({\partial{\varepsilon^{s}(k)}}/{\partial{k_y}})$ where 
	\begin{eqnarray}
		\frac{\partial{\varepsilon^{s}(k)}}{\partial{k_y}}={\frac{\hslash^2{k}(\gamma_{1}+\gamma_{2})}{m}}\sin{\theta}+s\left({\partial_{k}f}\sin{\theta}+\frac{\cos{\theta}{\partial_{\theta}f}}{k}\right).
	\end{eqnarray} 
	
	We perform some standard scaling, $\overline{k}=ka_{0}$, $\hbar\omega={\varepsilon_{H}}{\overline{\omega}}$. Thus:
	\begin{eqnarray}
		j^{c}_{x}={\frac{e}{\hbar}}{\frac{1}{a^2_{0}}}{a_0}{\varepsilon_{H}}[{e^2}{\frac{\hbar^2}{\varepsilon_{H}}}{\frac{1}{\hbar^2}}{a^{2}_0}]{\overline{j}}EE\\
		\frac{{e^3}{a_0}}{\hbar{\varepsilon_{H}}}{\overline{j}}EE=\frac{{e^3}{a_0}}{\hbar{\varepsilon_{H}}}{\overline{j}}{\frac{2{I_0}}{\epsilon_{0}{c}}}=\frac{{8{\pi}e}\alpha{a_0}}{{\varepsilon_{H}}}{\overline{j}}{I_0},
	\end{eqnarray}
	where $\varepsilon_{H}=13.6$ eV, $a_0=0.052$ nm are the energy of the ground state of the hydrogen atom and Bohr radius, respectively. Also, $j^{c}_{x}$ is in units of pA/m and $I_0$ is a unit of W/m$^2$ according to Ref. \cite{PhysRevB.95.035134}. We make use of $\epsilon_{0}=e^2/4\pi{\alpha}{\hbar{c}}$ with $\alpha={1}/{137}$. Therefore, $\sigma^{(2)}$ would be in units of pAm/W. Increasing the Fermi energy $\varepsilon^{0}_{F}$ and relaxation time $\tau$, the Fermi surface displacement increases so that it leads to a larger peak for photovoltaic effect current. In addition, there is a blue photovoltaic in the photovoltaic effect current when the Fermi energy increases. 
	Note that the second term in Eq. (\ref{eq:jx}) is dominant so that one can only consider this term and ignore other terms. Furthermore, as we mentioned before, we consider $\tau_1=\tau_2=\tau$.
	\begin{eqnarray}\label{eq:dominateterm}
		{\bf j}_{x,od}^c\approx\frac{e}{\hbar}\int \frac{kd{k}d{\theta}}{4\pi^2}({eE_0})^2\bigg\{\frac{{\cal R}^{*,ss'}_{k_x}{\Gamma}^{ss'}_{k_x}{\cal M}^{ss'}\left((\varepsilon^{s}-\varepsilon^{s'}-i\frac{\hbar}{\tau})^2+(\hbar{\omega})^2\right) (\varepsilon^{s}-\varepsilon^{s'})}{\left(\left(\varepsilon^{s}-\varepsilon^{s'}-i\frac{\hbar}{\tau}\right)^2-(\hbar{\omega})^2\right)^{2}\left(\varepsilon^{s}-\varepsilon^{s'}-i\frac{\hbar}{\tau}\right)}\bigg\}.
	\end{eqnarray}
	Assuming $\varepsilon^{s'}-\varepsilon^{s}=\hbar{\omega}$ and ignoring ${\hbar}^2/{\tau}^2$, the above equation is simplified as follows
	\begin{eqnarray}\label{eq:jappr}
		{\bf j}_{x,od}^c\approx\frac{e}{2\hbar}\int \frac{kd{k}d{\theta}}{4\pi^2}({eE_0})^2\bigg\{\frac{{\cal R}^{*,ss'}_{k_x}{\Gamma}^{ss'}_{k_x}{\cal M}^{ss'}}{\left(\varepsilon^{s'}-\varepsilon^{s}-\hbar{\omega}+i\frac{\hbar}{\tau}\right)^{2}}\bigg\}.
	\end{eqnarray}
	Here, the real part of photovoltaic current is as
	\begin{eqnarray}
		&&{\bf j}_{x,od}^c\approx\frac{e}{2\hbar}\int\frac{kd{k}d{\theta}}{4\pi^2}({eE_0})^2\frac{{\cal R}^{*,ss'}_{k_x}{\Gamma}^{ss'}_{k_x}{\cal M}^{ss'}(\tau=0)}{\left((\varepsilon^{s'}-\varepsilon^{s}-\hbar{\omega})^2+(\frac{\hbar}{\tau})^2\right)^{2}}\left[\left(\varepsilon^{s'}-\varepsilon^{s}-\hbar{\omega}\right)^2-(\frac{\hbar}{\tau})^2\right]\nonumber\\
		&-&\frac{e}{\hbar}\int\frac{kd{k}d{\theta}}{4\pi^2}({eE_0})^2\frac{{\vert{\cal R}^{ss'}_{k_x}\vert}^2(\frac{\hbar}{\tau})^2{\frac{\partial{\tau}}{\partial{k_x}}}}{\left((\varepsilon^{s'}-\varepsilon^{s}-\hbar{\omega})^2+(\frac{\hbar}{\tau})^2\right)^{2}}(\varepsilon^{s'}-\varepsilon^{s}-\hbar{\omega}).
	\end{eqnarray}
	As expected, there are two peaks in this figure so that the larger RPE current is a Lorentzian function centered around $\hbar{\omega}=\varepsilon^{s'}-\varepsilon^{s}$ which is due to the first part of the equation above. In order to find the position of the smaller peak which arises from the second part of this equation, we expand the denominator as  
	\begin{eqnarray}
		-\frac{e}{\hbar}\int\frac{kd{k}d{\theta}}{4\pi^2}({eE_0})^2{\frac{{\vert{\cal R}^{ss'}_{k_x}\vert}^2(\varepsilon^{s'}-\varepsilon^{s}-\hbar{\omega}){\frac{\partial{\tau}}{\partial{k_x}}}}{\left(2(\varepsilon^{s'}-\varepsilon^{s}-\hbar{\omega})^2+(\frac{\hbar}{\tau})^2\right)}}\Bigg\vert_{\omega\longrightarrow\omega-i\delta}\sim{\delta\left(\varepsilon^{s'}-\varepsilon^{s}-\hbar{\omega}-\frac{\hbar}{\tau}\right)},
	\end{eqnarray} 
	which shows the DC current with opposite sign occurs approximately around $\hbar{\omega}=\varepsilon^{s'}-\varepsilon^{s}-\frac{\hbar}{\tau}$. Using Eq. (\ref{eq:jappr}) and considering transition between band indexes $\lbrace{s=+1, s'=-1}\rbrace$ , the above equation takes the form
	\begin{eqnarray}
		{\bf j}_{x,od}^c (\varepsilon^{+}-\varepsilon^{-}\approx\hbar{\omega})\approx-\frac{e}{\hbar}({eE_0})^2{{{\frac{m}{{4\pi^2}\hbar^{2}(\gamma_{1}+\gamma_{2})}}}}\, {\rm Re} \, I_1(k_F),
	\end{eqnarray}
	where
	\begin{eqnarray}
		I_1(k_F)=\int{d{\theta}}{({\cal R}^{*,ss'}_{k_x}{\Gamma}^{ss'}_{k_x})_{k=k_F}{f(k_F, \theta)}}\frac{\left(2f(k_F, \theta)-\hbar{\omega}\right)^2-(\frac{\hbar}{\tau})^2}{\left[\left(2f(k_F, \theta)-\hbar{\omega}\right)^2+(\frac{\hbar}{\tau})^2\right]^2}\left(\frac{\partial\varepsilon^{+}(k)}{\partial{k_x}}-\frac{\partial\varepsilon^{-}(k)}{\partial{k_x}}+\frac{i\hbar}{{\tau^2}}\frac{\partial{\tau}}{\partial{k_x}}\right)_{{k=k_F}}.
	\end{eqnarray}
	Calculating the above integral, $I_1(k_F)$ can be written as
	\begin{eqnarray}
		I_1(k_F) = 382 \, {({\cal R}^{*,ss'}_{k_x}{\Gamma}^{ss'}_{k_x})_{k=k_F}{f(k_F, \theta=\pi)}}\left(\frac{\partial\varepsilon^{+}(k)}{\partial{k_x}}-\frac{\partial\varepsilon^{-}(k)}{\partial{k_x}}+\frac{i\hbar}{{\tau^2}}\frac{\partial{\tau}}{\partial{k_x}}\right)_{{k=k_F, \theta=\pi}}.
	\end{eqnarray}
	Based on the above approximations, the off-diagonal photovoltaic current at the resonant peak is determined by (note that only the real part is taken at the end) 
	\begin{eqnarray}\label{eq:approx}
		{\bf j}_{x,od}^c \propto {\frac{\omega\tau^2}{\hbar}}  \,\left(\frac{\partial{f(\varepsilon^{+}(k))}}{\partial{k_x}}-\frac{\partial{f(\varepsilon^{-}(k))}}{\partial{k_x}}+\frac{i\hbar}{\tau^2}\frac{\partial{\tau}}{\partial{k_x}}\right)_{k=k_F, \theta=\pi}.
	\end{eqnarray}

	\begin{figure}
		\includegraphics[scale=0.63]{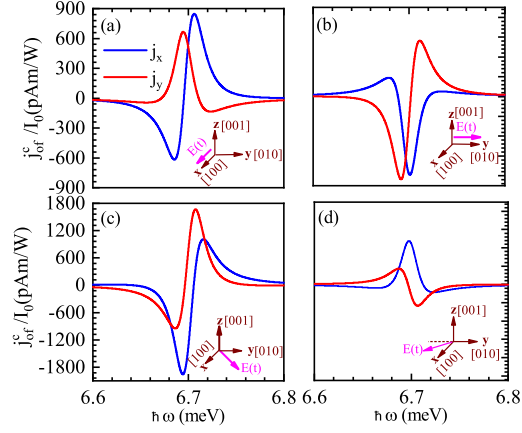}
		\caption{Resonant photovoltaic effect currents induced by light when the external time-dependent electric field is along four different crystal symmetry axes; $(a):[100], (b):[010], (c):  [110]$ and $(d):[1{\bar 1}0] $. The strength of the optical response depends strongly on the direction of the applied electric field.}  
		\label{Fig81}
	\end{figure}
	
	There is a similar trend for the peak photovoltaic current $j^{c}_y$ when the Fermi energy and the average relaxation time $\langle\tau(k_{F})\rangle$ increase. The maximum of the current $j^{c}_y$ curve fits very well with $(\langle\tau(k_F)\rangle)$ in medium and large relaxation time regions.

	Figure \ref{Fig81} shows the crystal symmetry dependence on the nonlinear optical response when the external electric field is along the crystal symmetry axis $[100], [010], [110]$ and $[1{\bar 1}0]$. As shown here, the strength of the optical response depends strongly on the direction of the applied electric field; $ j \propto |E_x {\hat i}+E_y {\hat j}|^2$. The reason lies in the fact that the ${\cal R}^{*,+-}_{k_i}{\Gamma}^{+-}_{k_i}{\cal M}^{+-}_{k_i}$ contribution depends to the direction in terms of the azimuth angle with $i$th being $x$ or $y$ as shown in Fig. \ref{Fig14}. Therefore, an important question is raised is there possible the optical transition response, $j^c_{od}$ vanishes in a specific direction? In general, for a given ${\bf E}=(E_x{\hat i}+E_y{\hat j})\cos \omega t$, our analytical calculations show that the optical transition response vanishes when the components of the electric field satisfy the following relation:
	\begin{eqnarray}\label{eq:e}
		&&\Gamma^{+-}_{{k_x}}{\cal M}^{*,+-}_{k_x} X^2++{\Gamma}^{*,+-}_{k_y}{\cal M}^{*,+-}_{k_y}\\
		&&+({ \Gamma}^{*,+-}_{k_y}{\cal M}^{*,+-}_{k_x}+{ \Gamma}^{*,+-}_{k_x}{\cal M}^{*,+-}_{k_y})X=0\nonumber
	\end{eqnarray}
	where $X=E_x/E_y$. 

	\begin{figure}
		\centering
		\includegraphics[scale=0.8]{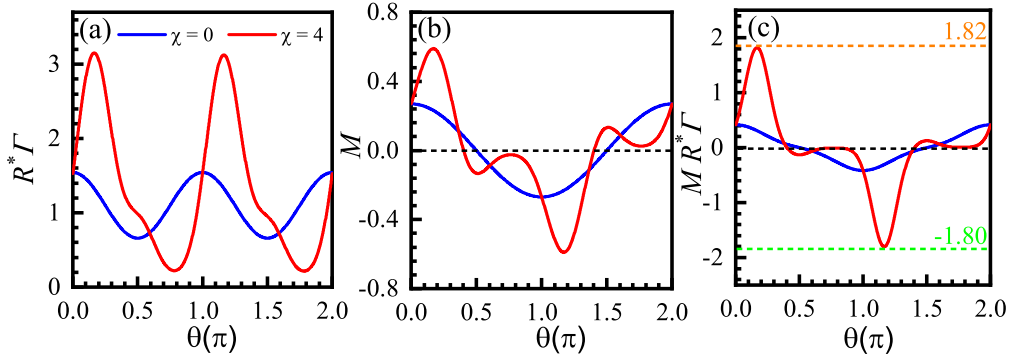}
		\caption{The sub-integral expressions in Eq. (\ref{eq:dominateterm}) for $\chi=0$ and $\chi=4$ including ${\cal R}^{*,ss'}_{k_x}{\Gamma}^{ss'}_{k_x}$, ${\cal M}$ and ${\cal R}^{*,ss'}_{k_x}{\Gamma}^{ss'}_{k_x}{\cal M}$ for a constant Fermi energy $\varepsilon^{0}_{F}=17.55$ meV.}
		\label{Fig9}
	\end{figure} 
	
	\begin{figure}
		\centering
		\includegraphics[scale=0.45]{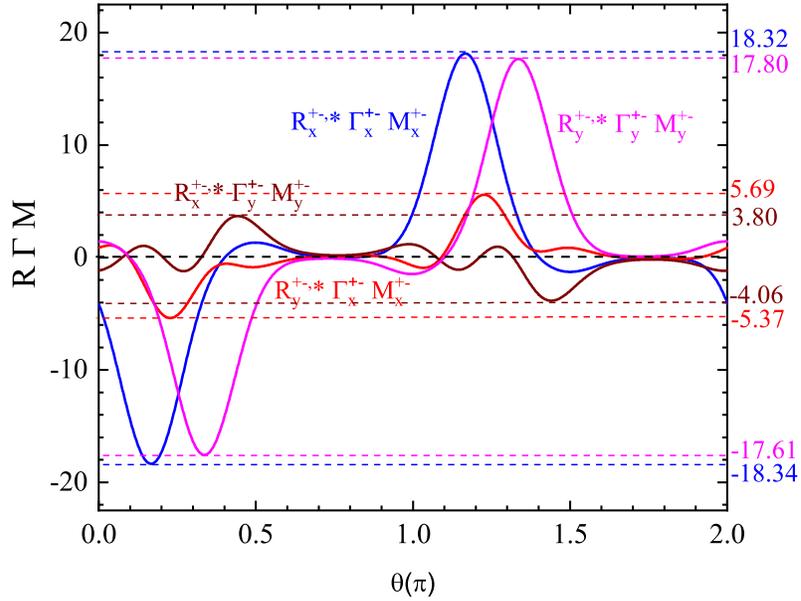}
		\caption{${\cal R}^{*,ss'}_{k_i}{\Gamma}^{ss'}_{k_i}{\cal M}_{k_i}$ along different directions with $i$th being $x$ and $y$ for a constant Fermi energy $\varepsilon^{0}_{F}=17.55$ meV. For given each electric field component, there are $j_{x od}$ and $j_{y, od}$. Terms proportional to ${\cal R}^{*,+-}_{i}$ refers to the case when $j_i$ is obtained where $i$ is $x$ or $y$. For each $j_{i, od}$, the electric field $E_{j}$ refers to ${\Gamma}^{+-}_{j}{\cal M}^{+-}_{j}$. Notice that ${\Gamma}^{+-}_{j}{\cal M}^{+-}_{j}$ defers along the $x$ and $y$ directions. }
		\label{Fig14}
	\end{figure} 
	
	In Fig. \ref{Fig9}, we show the sub-integral expressions in Eq. \ref{eq:dominateterm} including ${\cal R}^{*,ss'}_{k_x}{\Gamma}^{ss'}_{k_x}$, ${\cal M}$ and ${\cal R}^{*,ss'}_{k_x}{\Gamma}^{ss'}_{k_x}{\cal M}$ as a function of $\theta$ in the absence ($\chi=0$) and presence ($\chi=4$) of effective projected electric-dipole term. This figure demonstrates that for $\chi=0$ the result of integral in Eq. (\ref{eq:dominateterm}) becomes zero due to the existence of symmetry about $\theta$ while there is no such symmetry for $\chi\neq0$. Indeed, an oscillating electric field can be lead to electric-dipole transitions between heavy-hole and light hole states which it becomes vanish for $\chi=0$.
	
	\section{projection of the Hamiltonian matrix H onto the lowest
		subband}\label{project}
	The band Hamiltonian is as 
	\begin{equation}\label{eq:fullH}
		{\cal{H}}_{0}={\cal{H}}_{L}+V(z){I_{4}}+{\cal{H}}_{d_z},
	\end{equation}
	where $I_4$ is the $4\times 4$ identity matrix and
	\begin{equation}\label{eq:LuttingerHam}
		{\cal{H}}_L=\frac{\hbar^2}{2m}\left[\left(\gamma_1+\frac{5}{2}\gamma_2\right)k^2 I_4-2\gamma_2(\bm{k}\cdot\mathbf{J})^2\right]
	\end{equation}
	is the Luttinger Hamiltonian within the spherical approximation, with parameters $\gamma_1=6.85$ eV and $\gamma_2=$2.10 eV. 
	Here $I_{4}$ is the 4$\times$4 indentity matrix and $\textbf{J}=(J_{x},J_{y},J_{z})$ is the vector of spin-3/2 matrices so that its components are as 
	\begin{align}
		J_{x}={\frac{1}{2}}{\begin{bmatrix} 0 & \sqrt{3} & 0 & 0 \\ \sqrt{3} & 0 & 0 & 2 \\ 0 & 0 & 0 & \sqrt{3} \\ 0 & 2 & \sqrt{3} & 0\\ \end{bmatrix}}, J_{y}={\frac{i}{2}}{\begin{bmatrix} 0 &  -\sqrt{3}  & 0 & 0 \\ \sqrt{3} & 0 & 0 & -2 \\ 0 & 0 & 0 & \sqrt{3} \\ 0 & 2 & -\sqrt{3} & 0\\ \end{bmatrix}}, J_{z}={\frac{1}{2}}{\begin{bmatrix} 3 & 0 & 0 & 0 \\ 0 & 1 & 0 & 0 \\ 0 & 0 & -3 & 0 \\ 0 & 0 & 0 & -1\\ \end{bmatrix}}. 
	\end{align} 
	Also, ${\cal{H}}_{d_z}$ the effective projected electric-dipole Hamiltonian due to the presence of electrical field $\textbf{F}=F_{z}\hat{z}$ as follows:
	\begin{align}
		H_{d_z}={\frac{1}{\sqrt{3}}}e{a_B}{\chi}{F_z}{\lbrace{J_{x},J_{y}}\rbrace}, 
	\end{align}
	where $e{a_{B}}\simeq2.5$D ($a_{B}$ is the Bohr radius and D is a debye), $\chi$ is a parameter that controls the strength of the electric-dipole matrix elements. The Hamiltonian can be written as
	\begin{align}
		{\cal{H}}_{0}=\begin{bmatrix}
			{\mu}\left({\alpha_{+}}{k^2}+{\beta_{+}}{k^2_z}\right)+V(z) &  -\sqrt{3}{\mu}{\gamma_2}{k_{-}}{k_z} & 0 & -{\frac{\sqrt{3}}{2}}{\mu}{\gamma_2}{k^2_{-}}-i{e{a_B}{\chi}{F_z}} \\ -\sqrt{3}{\mu}{\gamma_2}{k_{+}}{k_z} & {\mu}\left({\alpha_{-}}{k^2}+{\beta_{-}}{k^2_z}\right)+V(z)  & -{\frac{\sqrt{3}}{2}}{\mu}{\gamma_2}{k^2_{-}}-i{e{a_B}{\chi}{F_z}} & 0\\ 0  & -{\frac{\sqrt{3}}{2}}{\mu}{\gamma_2}{k^2_{+}}+i{e{a_B}{\chi}{F_z}} & {\mu}\left({\alpha_{+}}{k^2}+{\beta_{+}}{k^2_z}\right)+V(z) &  \sqrt{3}{\mu}{\gamma_2}{k_{+}}{k_z}\\ -{\frac{\sqrt{3}}{2}}{\mu}{\gamma_2}{k^2_{+}}+i{e{a_B}{\chi}{F_z}}  & 0 & \sqrt{3}{\mu}{\gamma_2}{k_{-}}{k_z} & {\mu}\left({\alpha_{-}}{k^2}+{\beta_{-}}{k^2_z}\right)+V(z) \\
		\end{bmatrix},      
	\end{align}
	where $\mu={\frac{{\hslash}^2}{m}}$, $\alpha_{\pm}={\frac{\mu}{2}}\left(\gamma_{1}\pm\gamma_{2}\right)$, $\beta_{\pm}=\frac{{\gamma_{1}}\mp2{\gamma_{2}}}{2}$, $k_{\pm}=k_{x}\pm{i}k_{y}$ and $k^2=k^2_x+k^2_y$. Now, we project the Hamiltonian${\cal{H}}$ onto the lowest subband ($n=1$) to obtain a $4\times4$ Hamiltonian matrix. For this purpose, we consider states $\mid{HH\pm}\rangle=F^{1}_{HH\pm}(z)F(x)F(y)$ and $\mid{LH\pm}\rangle=F^{1}_{LH\pm}(z)F(x)F(y)$ so that F(x) and F(y) are plane waves. Considering $k_z=-{i}\frac{d}{dz}$ will have
	\begin{align}
		{\cal{H}}_{0}=\begin{bmatrix}
			\epsilon_{1} &  i\sqrt{3}{\mu}{\gamma_2}{k_{-}}{\eta_1} & 0 & -{\frac{\sqrt{3}}{2}}{\mu}{\gamma_2}{k^2_{-}}\xi-i{e{a_B}{\chi}{F_z}}\xi  \\ -i\sqrt{3}{\mu}{\gamma_2}{k_{+}}{\eta_1} & \epsilon_{2} & -{\frac{\sqrt{3}}{2}}{\mu}{\gamma_2}{k^2_{-}}\xi-i{e{a_B}{\chi}{F_z}}\xi & 0 \\ 0  & -{\frac{\sqrt{3}}{2}}{\mu}{\gamma_2}{k^2_{+}}\xi+i{e{a_B}{\chi}{F_z}}\xi & \epsilon_{1} & -\sqrt{3}i{\mu}{\gamma_2}{k_{+}}{\eta_1} \\ -{\frac{\sqrt{3}}{2}}{\mu}{\gamma_2}{k^2_{+}}\xi+i{e{a_B}{\chi}{F_z}}\xi & 0  & i\sqrt{3}{\mu}{\gamma_2}{k_{-}}{\eta_1} & \epsilon_{2} \\
		\end{bmatrix},      
	\end{align}
	where $\epsilon_{1(2)}=\epsilon_{H(L)}+(\gamma_{1}+\gamma_{1}){\frac{\hslash^2{k^2}}{2m}}$ and
	\begin{equation}\label{eq:xi}
		\xi = \int_{0}^{\infty} dzF^1_{\mathrm{HH}}(z)F^1_{\mathrm{LH}}(z),
	\end{equation}
	\begin{equation}\label{eq:eta}
		\eta_1 = \int_{0}^{\infty} dzF^1_{\mathrm{HH}}(z)\frac{d}{dz}F^1_{\mathrm{LH}}(z),
	\end{equation}
	Also, we define the following variables:
	\begin{equation}\label{eq:l}
		l = -\frac{\sqrt{3}\hbar^2}{2m}\xi \gamma_2
	\end{equation}
	\begin{equation}\label{eq:lambda}
		\lambda = \frac{\sqrt{3}\hbar^2}{m}\gamma_2\eta_1,
	\end{equation}
	so that ${\cal{H}}_{0}$ can be written as
	\begin{align}
		{\cal{H}}_{0}=\begin{bmatrix}
			\epsilon_{1} & i{\lambda}{k_{-}} & 0 & \ell{k^2_{-}}-i{e{a_B}{\chi}{F_z}}{\xi}\\ -i{\lambda}{k_{+}} & \epsilon_{2} & \ell{k^2_{-}}-i{e{a_B}{\chi}{F_z}}{\xi} & 0 \\ 0 & \ell{k^2_{+}}+i{e{a_B}{\chi}{F_z}}{\xi} &  \epsilon_{1} & -i{\lambda}{k_{+}} \\ \ell{k^2_{+}}+i{e{a_B}{\chi}{F_z}}{\xi}  & 0 & i{\lambda}{k_{-}}  & \epsilon_{2}\\
		\end{bmatrix}.     
	\end{align}
	Projecting the electric-dipole Hamiltonian $H_{d_x}$ onto the lowest subband, we obtain the following form:
	\begin{align}
		{\cal{H}}_{d_x}=\begin{bmatrix}
			0 & -i{e{a_B}{\chi}{E_x}}{\xi} & 0 & 0\\ i{e{a_B}{\chi}{E_x}}{\xi}  & 0 & 0 & 0 \\ 0 & 0 & 0 & -i{e{a_B}{\chi}{E_x}}{\xi} \\ 0 & 0 & i{e{a_B}{\chi}{E_x}}{\xi} & 0\\
		\end{bmatrix}.  
	\end{align}
	
	\section{Schrieffer-Wolff Transformation}\label{SWT}
	Schrieffer-Wolff (SW) transformation is a very important transformation in Quantum Many-Body Physics. SW transformation is a unitary transformation which removes the off-diagonal terms to the first order and hence serves as a way of diagonalization. We can choose the proper unitary operator which can either fully diagonalize the Hamiltonian or to some desired order. 
	\begin{equation}
		{\cal{H}}^{\prime}=U^{t}{{\cal{H}}}{U}
	\end{equation}
	\begin{equation}\label{eq:SHS}
		{\cal{H}}^{\prime}=e^{S}{{\cal{H}}}e^{-S}
	\end{equation}
	where $S$ is the generator of this transformation and is an anti-hermitian operator. Usually one requires this transformation to cancel the off-diagonal terms to the first order so that the following condition is satisfied \cite{PhysRevB.100.195117}.
	\begin{equation}
		[H_{0}, S]=-H_{2},
	\end{equation}
	where $H_0$, $H_1$, and $H_2$ are diagonal parts, off-diagonal parts of diagonal blocks and off-diagonal blocks of the hamiltonian ${\cal{H}}$, respectively. In order to apply SW transformation on Hamiltonian, we first omit the off-diagonal part of the diagonal block of ${\cal{H}}$ matrix via using the following rotation matrix:
	\begin{align}
		{\cal{R}}=\frac{1}{\sqrt{\epsilon_{+}+\epsilon_{-}}}\begin{bmatrix}
			\frac{i{\lambda}{k_{+}}}{\sqrt{\epsilon_{-}}} &  \frac{i{\lambda}{k_{+}}}{\sqrt{\epsilon_{+}}}  \\ \sqrt{\epsilon_{-}} & -\sqrt{\epsilon_{+}} \\
		\end{bmatrix}, 
	\end{align}
	where $\epsilon_{\pm}=\sqrt{(\Delta{\epsilon}/2)^2+{\lambda^2{k^2}}}\pm{\Delta{\epsilon}/2}$ so that $\Delta{\epsilon}=\epsilon_{2}-\epsilon_{1}$. Acting the rotation matrix on the full Hamiltonian matrix will be
	\begin{align}\label{eq:rotation1}
		{\cal{H}}=\begin{bmatrix}
			N & T\\ T^{\dagger} & S\\
		\end{bmatrix}\longrightarrow{\cal{H}}'=\begin{bmatrix}
			N' & T'\\ T'^{\dagger} & S'\\
		\end{bmatrix}, 
	\end{align}
	where $N'=N$,  
	\begin{align}
		{S'}={\cal{R}}^{\top}{\begin{bmatrix}
				\epsilon_{1} & -i{\lambda}{k_{+}}  \\  i{\lambda}{k_{-}} & \epsilon_{2}\\
		\end{bmatrix}}{\cal{R}}={\begin{bmatrix}
				\epsilon_{1}-\epsilon_{-} & 0 \\  0 & \epsilon_{1}+ \epsilon_{+} \\
		\end{bmatrix}}
	\end{align}
	and 
	\begin{align}\label{eq:rotation2}
		{T'}=T{\cal{R}}={\begin{bmatrix}
				0 & \ell{k^2_{-}}-i{e{a_B}{\chi}{F_z}}{\xi} \\  \ell{k^2_{-}}-i{e{a_B}{\chi}{F_z}}{\xi} &  0\\
		\end{bmatrix}}{\cal{R}}={\frac{1}{\sqrt{\epsilon_{+}+\epsilon_{-}}}}{\begin{bmatrix}
				{L_{1}}\sqrt{\epsilon_{-}} & -{L_{1}}\sqrt{\epsilon_{+}}  \\ i{\lambda}{k_{+}}(\frac{L_{1}}{\sqrt{\epsilon_{-}}}) &  i{\lambda}{k_{+}}(\frac{L_{1}}{\sqrt{\epsilon_{+}}}) \\
		\end{bmatrix}}
	\end{align}
	where $L_{1}=\ell{k^2_{-}}-ie{a_B}{\chi}{F_z}{\xi}$ and it is assumed $k\neq0$. Thus, $H_{0}$, $H_{1}$ and $H_{2}$ become as
	\begin{align}
		{H_{0}}=\begin{bmatrix}
			\epsilon_{1} & 0 & 0 & 0\\ 0 & \epsilon_{2} & 0 & 0\\ 0  & 0 & \epsilon_{1}-\epsilon_{-} & 0\\ 0  & 0 & 0 & \epsilon_{1}+\epsilon_{+} \\
		\end{bmatrix},
		{H_{1}}=\begin{bmatrix}
			0 & i{\lambda}{k_{-}} & 0 & 0\\ -i{\lambda}{k_{+}} & 0 & 0 & 0\\ 0  & 0 &  0 & 0\\ 0  & 0 & 0 & 0\\
		\end{bmatrix}
	\end{align}
	\begin{align}
		{H_{2}}={\frac{1}{\sqrt{\epsilon_{+}+\epsilon_{-}}}}\begin{bmatrix}
			0 & 0 & {L_{1}}\sqrt{\epsilon_{-}} & -{L_{1}}\sqrt{\epsilon_{+}} \\ 0  & 0 & i{\lambda}{k_{+}}(\frac{L_{1}}{\sqrt{\epsilon_{-}}}) &   i{\lambda}{k_{+}}(\frac{L_{1}}{\sqrt{\epsilon_{+}}}) \\ {L^{*}_{1}}\sqrt{\epsilon_{-}}  & -i{\lambda}{k_{-}}(\frac{L^{*}_{1}}{\sqrt{\epsilon_{-}}}) &  0 & 0\\ -{L^{*}_{1}}\sqrt{\epsilon_{+}}  & -i{\lambda}{k_{-}}(\frac{L^{*}_{1}}{\sqrt{\epsilon_{+}}}) & 0 & 0\\
		\end{bmatrix}.
	\end{align}
	We can also calculate $S$ matrix through the mentioned condition as
	\begin{align}
		S={\frac{1}{\sqrt{\epsilon_{+}+\epsilon_{-}}}}\begin{bmatrix}
			0 & 0 & -\frac{L_{1}}{\sqrt{\epsilon_{-}}}  & -\frac{L_{1}}{\sqrt{\epsilon_{+}}} \\ 0  & 0 & -\frac{i{\lambda}k_{+}L_{1}}{\sqrt{\epsilon_{-}}\epsilon_{+}}  &  \frac{i{\lambda}{k_{+}}{L_{1}}}{\sqrt{\epsilon_{+}}\epsilon_{-}}\\ \frac{L^{*}_{1}}{\sqrt{\epsilon_{-}}}   & \frac{-i{\lambda}{k_{-}}{L^{*}_{1}}}{\sqrt{\epsilon_{-}}\epsilon_{+}} &  0 & 0\\ \frac{L^{*}_{1}}{\sqrt{\epsilon_{+}}}  & \frac{i{\lambda}{k_{-}}{L^{*}_{1}}}{\epsilon_{-}\sqrt{\epsilon_{+}}} & 0 & 0\\
		\end{bmatrix},      
	\end{align} 
	Expanding the operator exponential using Eq. (\ref{eq:SHS}) one gets series expansion for the transformed Hamiltonian ${\cal{H}}_{eff}$ as
	\begin{equation}
		{\cal{H}}_{eff}=\sum^{\infty}_{j=0}{\frac{1}{(2j)!}}[H_0+H_1,S]^{(2j)}+\frac{1}{(2j+1)!}\sum^{\infty}_{j=0}[H_{2}, S]^{(2j+1)}.
	\end{equation}
	Since the off-diagonal term gets cancelled to the first order so the effective Hamiltonian to the second order is given by
	\begin{equation}
		{\cal{H}}_{eff}\approx{H_0}+{H_{1}}+\frac{1}{(2j+1)!}\sum^{2}_{j=0}[H_{2}, S]^{(2j+1)}.
	\end{equation}
	Notice that $[H_0+H_1,S]=0$ and using $S$ we will have
	\begin{equation}
		{\cal{H}}_{eff}=\begin{bmatrix}
			\epsilon_{1} & i{\lambda}{k_{-}} & 0 & 0\\ -i{\lambda}{k_{+}} & \epsilon_{2} & 0 &  0\\ 0 & 0 &  \epsilon_{1}-\epsilon_{-} & 0 \\ 0 & 0 & 0 & \epsilon_{1}+\epsilon_{+}\\
		\end{bmatrix}+C(k)\begin{bmatrix}
			0 & +i{\lambda}{k_{-}}{(\epsilon_{+}+\epsilon_{-})}  & 0 & 0  \\ -i{\lambda}{k_{+}}{(\epsilon_{+}+\epsilon_{-})}  &  0 & 0 & 0 \\ 0 & 0 &  2{\lambda^2}{k^2} & -\lambda{k}{\Delta{\epsilon}} \\ 0  & 0 & -\lambda{k}{\Delta{\epsilon}}  & -2{\lambda^2}{k^2}\\
		\end{bmatrix}.      
	\end{equation}
	where
	\begin{equation}
		C(k)=\sum^{2}_{j=0}\left(-\frac{{\vert{L_{1}}\vert}^2}{(\lambda{k})^2}\right)^{(j+1)}{\frac{2^{(2j+1)}}{{(2j+1)!}{(\epsilon_{+}+\epsilon_{-})}}}.
	\end{equation}
	If we consider only the LH1-HH1 subspace in above Hamiltonian, it gives the effective $2\times2$ Hamiltonian as
	\begin{equation}
		{\cal{H}}_{eff}=\begin{bmatrix}
			\epsilon_{1} & i{\lambda}{k_{-}}  \\ -i{\lambda}{k_{+}} & \epsilon_{2} \\\end{bmatrix}+C(k)\begin{bmatrix}
			0 & i{\lambda}{k_{-}}(\epsilon_{+}+\epsilon_{-}) \\ -i{\lambda}{k_{+}}(\epsilon_{+}+\epsilon_{-})   &  0 \\ 
		\end{bmatrix}.       
	\end{equation}
	Ignoring small phrases, we rewrite the above equation as
	\begin{eqnarray}
		{\cal{H}}_{{\text eff}}=\begin{bmatrix}
			\epsilon_{1} & 0 \\ 0 & \epsilon_{2} \\\end{bmatrix}+i\left({\lambda}-\frac{4e{a_B}{\chi}{F_z}{\xi}{\ell}}{\lambda}{\sin{2\theta}}\right)\begin{bmatrix}
			0 & {k_{-}} \\ -{k_{+}}  & 0 \\ \end{bmatrix}-2i\left({\frac{\ell^2}{\lambda}}-8\frac{e{a_B}{\chi}{F_z}{\xi}{\ell}^3}{3{\lambda}^3}{\sin{2\theta}}\right){k^3}\begin{bmatrix} 0 & e^{-i\theta}  \\ -e^{i\theta}  & 0 \\ \end{bmatrix}\nonumber\\
		+\frac{4i{\ell^4}}{3{\lambda}^3}\left(1-\frac{6e{a_B}{\chi}{F_z}{\xi}{\ell}}{5{\lambda}^2}{\sin{2\theta}}\right){k^5}\begin{bmatrix} 0 & e^{-i\theta}  \\ -e^{i\theta}  & 0 \\ \end{bmatrix}-{\frac{4i{\ell}^6}{15{\lambda}^5}}{k^7}\begin{bmatrix} 0 & e^{-i\theta}  \\ -e^{i\theta}  & 0 \\ \end{bmatrix},     
	\end{eqnarray}
	\begin{figure}[ht]
		\centering
		\includegraphics[scale=0.6]{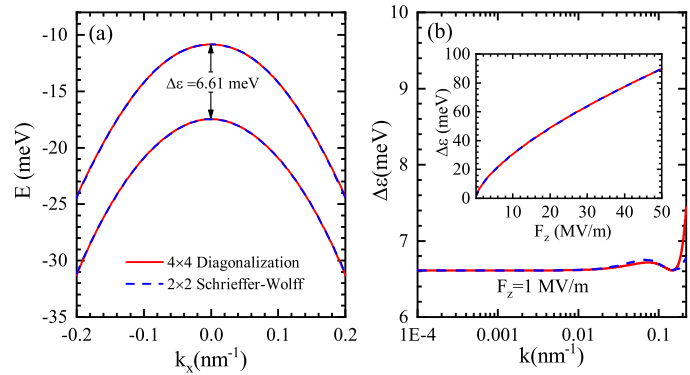}
		\caption{(a) 
			Light and heavy hole bands of GaAs (b) heavy and light hole splitting computed by diagonalizing the effective $2\times2$ Hamiltonian obtained from the SW transformation, as described in the main text (in blue), by diagonalizing the $4\times4$ Hamiltonian in the lowest valence-band subband (in red) for $F_z=1\,MV/m$. The inset indicates the band gap between the light and heavy hole states as a function of the electric field $F_z$.}
		\label{Fig10}
	\end{figure}
	\begin{figure}[ht]
		\centering
		\includegraphics[scale=0.6]{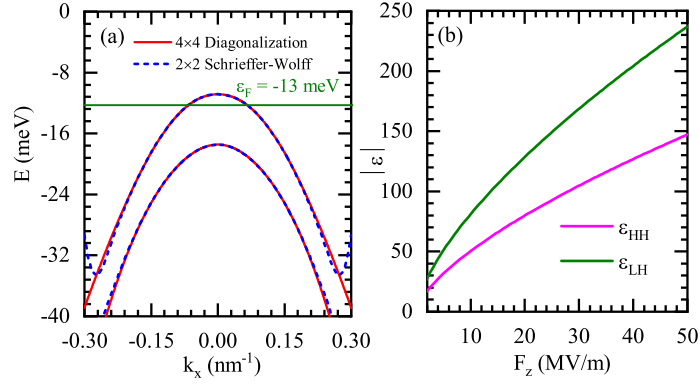}
		\caption{(a) Light and heavy hole bands of GaAs in the interval $-0.3\,{nm^{-1}}<k_x<0.3\,{nm^{-1}}$ for above figure (b) The light and heavy hole eigenenergies as a function of the electrical field.}
		\label{Fig11}
	\end{figure}
	We simplify the effective Hamiltonian and express it in form
	\begin{eqnarray}
		{\cal{H}}_{eff}=\left(\frac{\epsilon_{1}+\epsilon_{2}}{2}\right){\sigma_{0}}-\frac{\Delta{\epsilon}}{2}\sigma_{z}+i(\lambda+{\beta_{\chi{1}}}{\sin{2\theta}})\left(k_{-}{\sigma_{+}}-k_{+}{\sigma_{-}}\right)+i(\gamma_{R1}+\beta_{\chi{2}}{\sin{2\theta}}){k^3}\left(e^{-i\theta}{\sigma_{+}}-e^{i\theta}{\sigma_{-}}\right)\nonumber\\
		+i(\gamma_{R2}+\beta_{\chi{3}}{\sin{2\theta}}){k^5}\left(e^{-i\theta}{\sigma_{+}}-e^{i\theta}{\sigma_{-}}\right)+i{\gamma_{R3}}{k^7}\left(e^{-i\theta}{\sigma_{+}}-e^{i\theta}{\sigma_{-}}\right).    
	\end{eqnarray}
	Here, we defined ${\beta_{\chi{1}}}=-4e{a_B}{\chi}{F_z}{\xi}{\ell}/\lambda$, ${\beta_{\chi{2}}}=16e{a_B}{\chi}{F_z}{\xi}{\ell^3}/{3\lambda^3}$, ${\beta_{\chi{3}}}=-24e{a_B}{\chi}{F_z}{\xi}{\ell^5}/{15\lambda^5}$,${\gamma_{R1}}=-2{\ell}^2/{\lambda}$, ${\gamma_{R2}}=4{\ell}^4/{3\lambda^3}$  and ${\gamma_{R3}}=-4{\ell}^6/{15\lambda^5}$.
	We calculate the dispersion relation for the above $2\times2$ Hamiltonian via the diagonalization process as can be observed in Fig.~\ref{Fig10} (blue dashed line). To verify the effective two-dimensional Hamiltonian ${\cal{H}}_{\text {eff}}$ obtained from perturbation theory (SW transformation), we have also numerically diagonalized the four-dimensional Hamiltonian, resulting in the dispersion relation shown in the red line of Fig.~\ref{Fig10}(a). As can be observed in Fig. \ref{Fig10}(a), there is a gap $\Delta=6.61$ meV between the heavy-hole/light-hole states for $F_z=1$ MV/m and band dispersions derived from two method are consistent for $k\lesssim{0.2}$ nm$^{-1}$. We also investigate the value of heavy and light hole splitting obtained from diagonalizing $2\times2$ Hamiltonian based on SW transformation and diagonalizing $4\times4$ Hamiltonian as shown in Fig. \ref{Fig10}(b). In this figure, SW transformation break downs at large $k$ ($k\gtrsim 0.2$ nm$^{-1}$). For a two-dimensional hole gas with Fermi wavevector $k_F=k$, this implies a low sheet density, $n_p=k^2/(2\pi)\lesssim {6.4\times10^{11}}$ cm$^{-2}$. When the electrical field $F_z$ increases, the bandgap between the heavy hole/light hole states increases as can be observed in the inset of Fig. \ref{Fig10}(b). It is evident that the compatibility range of wave vectors for the values of splitting obtained from two methods in the presence of a larger electrical field increases so that it reaches $k=0.3$ nm$^{-1}$ which implies a hole density of $n_p= 14.3\times10^{11}$ cm$^{-2}$.    
	\begin{table}
		\caption{The calculated values of GaAs valence-band parameters.}  
		\begin{ruledtabular}
			\begin{center}
				\begin{tabular}{ c c c c c c c c c c c}
					$F_z$ & $\Delta{\epsilon}$ & $\lambda$ & ${\gamma_{R1}}$ & ${\gamma_{R2}}$ & ${\gamma_{R3}}$ & ${\beta_{\chi1}}$ & ${\beta_{\chi2}}$ & ${\beta_{\chi3}}$ \\
					(MV$m^{-1}$) & (meV) & (meV nm) & (meV nm$^{3}$) & (meV nm$^{5}$)  & (meV nm$^{7}$) & (meV nm)  & (meV nm$^{3}$) & (meV nm$^{5}$)\\
					\hline
					$1$ & $6.61$ & $15.99$  & $-1.457\times10^{3}$ & $4.42\times10^{4}$ & -$4.02\times10^{5}$ & $0.22$ & -$13.51$ & $184.67$ \\
					$10$ & $30.68$ & $29.89$ & $-966$ & $1.04\times10^{4}$ & $-3.36\times10^{4}$ & $1.47$ & -$31.81$ & $154.22$   
				\end{tabular}
			\end{center}
		\end{ruledtabular}
	\end{table}
	
	As can be seen in Fig.~\ref{Fig13}, the Rashba spin-orbit coupling coefficients vary as a function of electric field $F_z$. These coefficients are independent of the wave vector as a result of the SW transformation. As seen in this figure, the coefficients $\lambda$ and $\beta_{\chi1}$ increase linearly with the electric field. The other Rashba spin-orbit coupling coefficients at low electric field increase with $F_z$ which is in agreement with the trends reported in other papers \cite{papadakis1999effect} and then saturate at a larger electric field or decrease with increasing electric field $F_z$ which is in accordance with the experimental findings \cite{habib2004negative}.
	\begin{figure}
		\centering
		\includegraphics[scale=0.5]{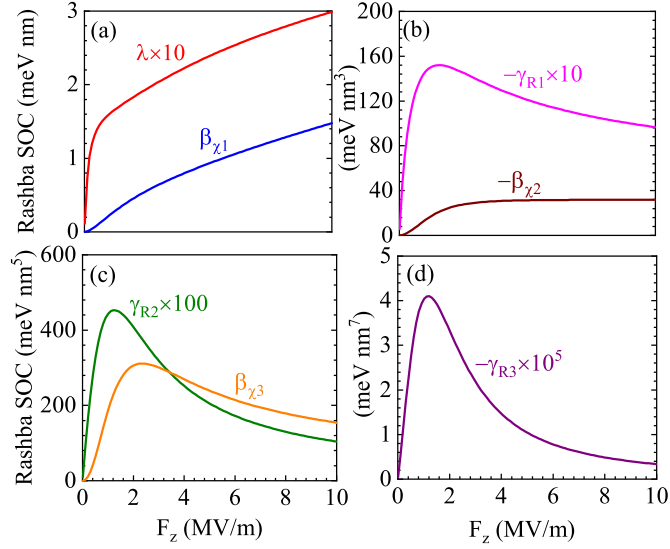}
		\caption{Magnitude of the Rashba spin-orbit coupling coefficients as a function of the electric field, $F_z$.}
		\label{Fig13}
	\end{figure}
	We also apply the SW transformation by using $S$ matrix calculated on the electric-dipole Hamiltonian ${\cal{H}}_{d_x}$. For this purpose, we first omit the off-diagonal part of the diagonal block of ${\cal{H}}_{d_x}$ matrix by using the following rotation matrix:
	\begin{align}
		{\cal{R}}=\frac{1}{\sqrt{2}}\begin{bmatrix}
			i &  -i  \\ 1 & 1 \\
		\end{bmatrix}. 
	\end{align}
	Using this rotation matrix [see Eqs. (\ref{eq:rotation1})-(\ref{eq:rotation2})], the Hamiltonian ${\cal{H}}_{d_x}$ can be written as
	\begin{align}
		{\cal{H}}_{d_x}=\begin{bmatrix}
			0 & -i{e{a_B}{\chi}{E_x}}{\xi} & 0 & 0\\ i{e{a_B}{\chi}{E_x}}{\xi}  & 0 & 0 & 0 \\ 0 & 0 & -{e{a_B}{\chi}{E_x}}{\xi} & 0  \\ 0 & 0 & 0 & {e{a_B}{\chi}{E_x}}{\xi}\\
		\end{bmatrix}.  
	\end{align}
	Then, the transformed Hamiltonian ${\cal{H}}^{'}_{d_x}$ to the second order is calculated as
	\begin{equation}
		{\cal{H}}^{'}_{d_x}\approx{H^{'}_{0}}+{H^{'}_{1}}+\frac{1}{2}\left[[H^{'}_{0}+H^{'}_{1}, S], S\right],
	\end{equation}
	where 
	\begin{align}
		{H^{'}_{0}}=\begin{bmatrix}
			0 & 0 & 0 & 0\\ 0 & 0 & 0 & 0\\ 0  & 0 & -{e{a_B}{\chi}{E_x}}{\xi} & 0\\ 0  & 0 & 0 & {e{a_B}{\chi}{E_x}}{\xi} \\
		\end{bmatrix},
		{H^{'}_{1}}=\begin{bmatrix}
			0 & -i{e{a_B}{\chi}{E_x}}{\xi} & 0 & 0\\ i{e{a_B}{\chi}{E_x}}{\xi} & 0  & 0 & 0\\ 0  & 0 &  0 & 0\\ 0  & 0 & 0 & 0\\
		\end{bmatrix}.
	\end{align}
	Using $S$ matrix, the transformed Hamiltonian takes following form:
	\begin{equation}
		{\cal{H}}^{'}_{d_x}=\begin{bmatrix}
			0 & -\kappa & 0 & 0\\ \kappa & 0 & 0 &  0\\ 0 & 0 &  -\kappa & 0 \\ 0 & 0 & 0 & \kappa\\
		\end{bmatrix}\nonumber\\
		+\kappa_0\begin{bmatrix}
			-\Delta{\varepsilon} & i(\epsilon_{+}+\epsilon_{-})+2i{\lambda}{k_{-}} & 0 & 0  \\ -i(\epsilon_{+}+\epsilon_{-})-2i{\lambda}{k_{+}} &  \Delta{\varepsilon} & 0 & 0 \\ 0 & 0 & (\epsilon_{+}+\epsilon_{-})+2{\lambda}k{\cos{\theta}} & (\epsilon_{-}e^{-i\theta}-\epsilon_{+}e^{i\theta}) \\ 0  & 0 & (\epsilon_{-}e^{i\theta}-\epsilon_{+}e^{-i\theta})  &  -(\epsilon_{+}+\epsilon_{-})-2{\lambda}k{\cos{\theta}} \\
		\end{bmatrix}.   
	\end{equation}
	where $\kappa_0=\frac{{\vert{L_1}\vert}^2{e{a_B}{\chi}{E_x}}{\xi}}{({\lambda}k)^2(\epsilon_{+}+\epsilon_{-})}$ and $\kappa=i{e{a_B}{\chi}{E_x}}{\xi}$. If we consider only the LH1-HH1 subspace in above Hamiltonian, it gives the effective $2\times2$ Hamiltonian as
	\begin{equation}
		{\cal{H}}^{'}_{d_x}=\begin{bmatrix}
			0 & -\kappa \\ \kappa  & 0 \\\end{bmatrix}+\kappa_0\begin{bmatrix}
			-\Delta{\varepsilon} & i(\epsilon_{+}+\epsilon_{-})+2i{\lambda}{k_{-}}  \\ -i(\epsilon_{+}+\epsilon_{-})-2i{\lambda}{k_{+}} & \Delta{\varepsilon}\\ \end{bmatrix}.
	\end{equation}
	Ignoring small terms, it is simplified as
	\begin{equation}
		{\cal{H}}^{'}_{d_x}=\frac{1}{2}{e{a_B}{\chi}{\xi}{E_x}}\begin{bmatrix}
			-\beta_{k}{\Delta{\varepsilon}}k^2 & 2i\left(\alpha{k^2}+\beta_{k}{\lambda}k^3{e^{-i\theta}}-1\right) \\ -2i\left(\alpha{k^2}+\beta_{k}{\lambda}k^3{e^{i\theta}}-1\right) & \beta_{k}{\Delta{\varepsilon}}k^2 \\
		\end{bmatrix},
	\end{equation}
	where $\alpha=(\ell/\lambda)^2$ and $\beta_{k}=\alpha/\sqrt{({\Delta{\varepsilon}}/2)^2+(\lambda{k})^2}$. Then, we set the coordinates of above Hamiltonian by a rotation matrix formed from the normalized eigenvectors of ${\cal H}_{eff}$ [see Eq.~\ref{eq:eigenvectors}] as
	\begin{equation}
		{\cal{R}}=\frac{1}{\sqrt{2f({\bm k})}}\begin{bmatrix}
			\sqrt{f({\bm k})-\frac{\Delta{\varepsilon}}{2}} & -\sqrt{f({\bm k})+\frac{\Delta\varepsilon}{2}}  \\ -i\sqrt{f({\bm k})+\frac{\Delta\varepsilon}{2}}{e^{i\theta}}  &  -i\sqrt{f({\bm k})-\frac{\Delta\varepsilon}{2}}{e^{i\theta}} \\
		\end{bmatrix}.
	\end{equation}
	Acting the rotation matrix on the Hamiltonian will be
	\begin{equation}
		{\cal H}''_{d_x}={\cal{R}}^{\dagger}{{\cal H}'_{d_x}}{\cal{R}}={e{a_B}{\chi}{\xi}{E_x}}\begin{bmatrix}
			\Pi({\bm k}) & \zeta_{\bm k}+(1-\alpha{k^2})\Omega_{\bm k}  \\ \zeta^{*}_{\bm k}+(1-\alpha{k^2})\Omega^{*}_{\bm k}  & -\Pi({\bm k}), \\
		\end{bmatrix}
	\end{equation}
	with
	\begin{eqnarray}
		\Pi({\bm k})=\frac{\beta_{k}(\Delta{\varepsilon}k)^2}{4f({\bm k})}+\frac{\sqrt{f({\bm k})^2-(\frac{\Delta{\varepsilon}}{2})^2}}{f({\bm k})}\Bigg(\beta_{k}{\lambda}k^3-(1-\alpha{k^2}){\cos{\theta}}\Bigg),\nonumber\\
		\zeta_{\bm k}=\frac{\beta_{k}{\Delta{\varepsilon}}k^2}{2f({\bm k})}\left(\sqrt{f({\bm k})^2-(\frac{\Delta{\varepsilon}}{2})^2}-\lambda{k}\right),\quad\Omega_{\bm k}=\frac{\Delta{\varepsilon}\cos{\theta}}{2f({\bm k})}+i{\sin{\theta}},
	\end{eqnarray}
	where $f({\bm k})=\varepsilon^{+}_{\bm k}-\varepsilon_{0}$ [see Eqs.~\ref{eq:eigenvalues} and \ref{eq:f(k)}]. The bove Hamiltonian for the electrical field $E_y$ after performing the calculations is written as
	\begin{equation}
		{\cal H}''_{d_y}={e{a_B}{\chi}{\xi}{E_x}}\begin{bmatrix}
			\Pi_{y}({\bm k}) & \zeta_{\bm k}-(1-\alpha{k^2})\Omega^{y}_{\bm k}  \\ \zeta^{*}_{\bm k}-(1-\alpha{k^2})\Omega^{*,y}_{\bm k}  & -\Pi_{y}({\bm k}), \\
		\end{bmatrix}
	\end{equation}
	with
	\begin{eqnarray}
		\Pi_{y}({\bm k})=\frac{\beta_{k}(\Delta{\varepsilon}k)^2}{4f({\bm k})}+\frac{\sqrt{f({\bm k})^2-(\frac{\Delta{\varepsilon}}{2})^2}}{f({\bm k})}\Bigg(\beta_{k}{\lambda}k^3+(1-\alpha{k^2}){\sin{\theta}}\Bigg),\quad\Omega^{y}_{\bm k}=\frac{\Delta{\varepsilon}\sin{\theta}}{2f({\bm k})}+i{\cos{\theta}}
	\end{eqnarray}

	\section{Solution to triangular confinig potential}\label{Solution}
	The envelope function $F^n_{\nu}(z)$ is obtained by solving the following differential equation:
	\begin{equation}
		\left[{-\frac{\hbar^2}{2{m_\nu}}}{\frac{d^2}{d{z^2}}}+{e{F_z}z}\right]{F^n_{\nu}(z)}={\epsilon^{n}_{\nu}}{F^n_{\nu}(z)} 
	\end{equation}
	Using the change of variable $z=\rho{x}$, the above equation is rewritten as 
	\begin{equation}
		\left[{-\frac{\hbar^2}{2{m_\nu}e{F_z}{\rho}^3}}{\frac{d^2}{d{x^2}}}+x\right]{F^n_{\nu}({x})}=\frac{{\epsilon^{n}_{\nu}}}{\rho{e}{F_z}}{F^n_{\nu}({x})} 
	\end{equation}
	Also, we use the definitions $\rho=({\hbar^2}/{2{m_\nu}e{F_z}})^{1/3}$ and $a_n=\frac{{\epsilon^{n}_{\nu}}}{\rho{e}{F_z}}$ in this equation:
	\begin{equation}
		{{F''}^n_{\nu}({x})}={(x-a_n)}{F^n_{\nu}({x})}. 
	\end{equation}
	The solutions of the previous equation are two linearly independent Airy functions Ai$(x-a_n)$ and Bi$(x-a_n)$. The Bi$(x-a_n)$ solution diverges for considerable positive argument and does not satisfy the boundary condition ${F^n_{\nu}({z\rightarrow\infty})}$, therefore it is excluded. The energy eigenvalues are determined by the boundary condition imposed by the infinite wall at the origin, namely that $F^n_{\nu}(z=0)=$Ai$(-a_n)=0$, so $a_n$ is the nth zero of Ai$(-z)$. Ultimately, the envelopes are expressed via the Airy functions as follows:
	\begin{equation} 
		F^n_{\nu}(z)={C_v}{Ai(\frac{z}{\rho}-a_n)},
	\end{equation}  
	where ${C_v}$ is the normalization factor which can be determined from the wave function normalization condition as
	\begin{equation} 
		{{C_v}^2}\int^{\infty}_{0}{{Ai}^2(\frac{z}{\rho}-a_n)}{dz}=1.
	\end{equation} 
	Using the change of variable $\Lambda(z)=z/{\rho}-a_n$, the above equation is rewritten as
	\begin{equation} 
		{\rho}{{C_v}^2}\int^{+\infty}_{\Lambda(0)}{{Ai}^2(\Lambda(z))}{d{\Lambda(z)}}=1.
	\end{equation} 
	If we employ the expression \cite{PhysRevB.5.4891}
	\begin{equation}
		\int^{\infty}_{z}{Ai}^2(x){dx}=-{z}{Ai}^2(z)+{Ai'}^2(z),
	\end{equation}
	the normalization factor $C_v$ is written as
	\begin{equation}
		C_v=\left(\frac{(\frac{2{m_\nu}e{F_z}}{\hbar^2})^{1/3}}{{Ai'}^2(\Lambda(0))-{{\Lambda(0)}{Ai}^2(\Lambda(0))}}\right)^{1/2}.
	\end{equation} 
	In summary, the envelope functions are expressed as
	\begin{equation} 
		F^n_{\nu}(z)={C_v}Ai\left((\frac{2{m_\nu}e{F_z}}{\hbar^2})^{1/3}z-a_n\right),
	\end{equation}  
	and the eigenenergies are given by
	\begin{equation}
		a_n=\frac{{\epsilon^{n}_{\nu}}}{\rho{e}{F_z}}\Longrightarrow{\epsilon^{n}_{\nu}}=\left(\frac{e^2{\hbar^2}{F^2_z}}{2{m_\nu}}\right)^{1/3}{a_n}.  
	\end{equation} 
	\section{The relaxation time for a short-range disorder}\label{RT}
	Here we derive an approximation for the relaxation time. As mentioned, $\delta$ function has given as 
	\begin{eqnarray}
		\delta(\varepsilon^{\pm}(k')-\varepsilon^{\pm}(k))={\delta}\left({\frac{\hslash^{2}}{2m}}{(\gamma_{1}+\gamma_{2})}({k'}^2-k^2)\pm{f(k', \theta')}\mp{f(k, \theta)}\right),
	\end{eqnarray}
	We expand $\delta$ function up to the second order as
	\begin{eqnarray}
		\delta(\varepsilon^{\pm}(k')-\varepsilon^{\pm}(k))\approx\delta(\varepsilon'_{0}-\varepsilon_{0})\pm\left(f(k', \theta')-f(k, \theta)\right)\frac{\partial}{\partial{\varepsilon'_{0}}}{\delta(\varepsilon'_{0}-\varepsilon_{0})}+{\left(f(k',\theta')-f(k, \theta)\right)^2}\frac{\partial^{2}}{\partial{\varepsilon'^{2}_{0}}}{\delta(\varepsilon'_{0}-\varepsilon_{0})}, 
	\end{eqnarray}
	where $\varepsilon^{(')}_{0}=A{k^{(')}}^2$ with $A={\frac{\hslash^{2}}{2m}}(\gamma_{1}+\gamma_{2})$. Changing the partial derivatives with respect to $k'$ will have
	\begin{eqnarray}
		\frac{\partial}{\partial{\varepsilon'_{0}}}=\frac{1}{2{k'}{A}}{\frac{\partial}{\partial{k'}}},\nonumber
	\end{eqnarray}
	\begin{eqnarray}
		\frac{\partial^{2}}{\partial{\varepsilon'^{2}_{0}}}=\frac{-1}{4{A^2}k'^3}{\frac{\partial}{\partial{k'}}}+\frac{1}{4{A^2}k'^2}{\frac{\partial^{2}}{\partial{k'}^2}}.
	\end{eqnarray}
	Now, we make use of an equality for which $f(x){\delta^{(n)}(x)}={(-1)^{n}}{f^{(n)}(x)}{\delta(x)}$. Therefore
	\begin{eqnarray}
		\delta(\varepsilon^{\pm}(k')-\varepsilon^{\pm}(k))=(1\pm{\frac{B}{2A{k'}^2}}\mp{\frac{\partial_{k'}f}{2A{k'}}}-{\frac{3B{\partial_{k'}f}}{2A^2{k'}^3}}+{\frac{3B^2}{4A^2{k'}^4}}+\frac{1}{2}(\frac{\partial_{k'}f}{A{k'}})^2+\frac{B\partial^2_{k'}f}{2A^2{k'}^2}){{\delta(\varepsilon'_{0}-\varepsilon_{0})}},
	\end{eqnarray}
	where $B=f(k', \theta')-f(k, \theta)$. In addition, using the properties of delta function, will have
	\begin{eqnarray}
		{\delta(\varepsilon'_{0}-\varepsilon_{0})}={\frac{1}{2kA}}\left(\delta(k'-k)+\delta(k'+k)\right) 
	\end{eqnarray}  
	so the following integral yield as
	\begin{eqnarray}
		\int_{0}^{+\infty}{k'{dk'}}{\delta(\varepsilon^{\pm}(k')-\varepsilon^{\pm}(k))}=\frac{1}{2A}\left(1\pm{\frac{B}{2A{k}^2}}\mp{\frac{\partial_{k}f}{2A{k}}}-{\frac{3B{\partial_{k}f}}{2A^2{k}^3}}+{\frac{3B^2}{4A^2{k}^4}}+\frac{1}{2}(\frac{\partial_{k}f}{A{k}})^2+\frac{B\partial^2_{k}f}{2A^2{k}^2}\right),
	\end{eqnarray}  
	where $B=f(k, \theta')-f(k, \theta)$.
	\begin{figure}
		\centering
		\includegraphics[scale=0.7]{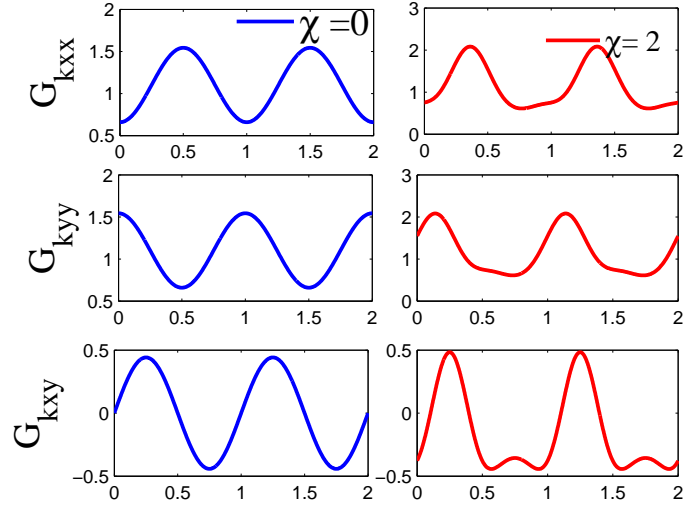}
		\caption{The independent contributions of quantum metric}
		\label{metric}
	\end{figure}
	\section{Quantum metric}\label{QM}
	
	We obtain the quantum metric of the system defined as one of the quantum geometric quantities. For an external time-dependent
	homogeneous electric field given by $\textbf{E}(t)=\textbf{E}_0{e^{-i\omega{t}}}$ with $\textbf{E}_0={\lbrace}E^{x}_0, E^{y}_0\rbrace$, the current depends on the independent contributions of quantum metric $\cal{G}$ including ${\cal G}^{ss'}_{k_{yy}}=\vert{R^{ss'}_{k_{y}}}\vert^2$, ${\cal G}^{ss'}_{k_{xx}}=\vert{R^{ss'}_{k_{x}}}\vert^2$ and ${\cal G}^{ss'}_{k_{xy}}=({1/2})\left({R^{s's}_{k_{x}}R^{ss'}_{k_{y}}+R^{ss'}_{k_{x}}R^{s's}_{k_{y}}}\right)$. Note that the quantity of the quantum geometric is symmetric (e.g, ${\cal G}^{ss'}_{k_{xy}}={\cal G}^{ss'}_{k_{yx}}$) and the condition ${\cal G}^{ss'}={\cal G}^{s's}$ is satisfied. For the different band indices (i.e, $s=+$, $s'=-$), these contributions can be written as
	\begin{eqnarray}
		{\cal G}^{+-}_{k_{xx}}&&=(\frac{\Delta{\varepsilon}}{4f^2})^2\left(({\lambda}'+3{\gamma}'_{1}k^2+5{\gamma}'_{2}k^4+7{\gamma}_{R3}k^6)\cos{\theta} -2\sin{\theta}\cos{2\theta}(\beta_{\chi1}+\beta_{\chi2}k^2+\beta_{\chi3}k^4)\right)^2\nonumber\\
		&&+(\frac{\sin{\theta}}{2f})^2({\lambda}'+{\gamma}'_{1}k^2+{\gamma}'_{2}k^4+{\gamma}_{R3}k^6)^2  
	\end{eqnarray} 
	\begin{eqnarray}
		{\cal G}^{+-}_{k_{yy}}&&=(\frac{\Delta{\varepsilon}}{4f^2})^2\left(({\lambda}'+3{\gamma}'_{1}k^2+5{\gamma}'_{2}k^4+7{\gamma}_{R3}k^6)\sin{\theta}+2\cos{\theta}\cos{2\theta}(\beta_{\chi1}+\beta_{\chi2}k^2+\beta_{\chi3}k^4)\right)^2\nonumber\\
		&&+(\frac{\cos{\theta}}{2f})^2({\lambda}'+{\gamma}'_{1}k^2+{\gamma}'_{2}k^4+{\gamma}_{R3}k^6)^2  
	\end{eqnarray}
	
	\begin{eqnarray}
		{\cal G}^{+-}_{k_{xy}}&&=\frac{1}{4f^2}\left((\frac{\Delta{\varepsilon}}{2f})^2({\lambda}'+3{\gamma}'_{1}k^2+5{\gamma}'_{2}k^4+7{\gamma}_{R3}k^6)^2-({\lambda}'+{\gamma}'_{1}k^2+{\gamma}'_{2}k^4+{\gamma}_{R3}k^6)^2\right)\cos{\theta}\sin{\theta}\nonumber\\
		&&+\frac{1}{2}(\frac{\Delta{\varepsilon}}{2f^2})^2({\lambda}'+3{\gamma}'_{1}k^2+5{\gamma}'_{2}k^4+7{\gamma}_{R3}k^6)(\beta_{\chi1}+\beta_{\chi2}k^2+\beta_{\chi3}k^4)\cos^2{2\theta}\nonumber\\
		&&-(\frac{\Delta{\varepsilon}}{2f^2})^2(\beta_{\chi1}+\beta_{\chi2}k^2+\beta_{\chi3}k^4)^2\sin{4\theta}\cos{2\theta}   
	\end{eqnarray}
	

\end{widetext}
\bibliography{ref-GaAs2}
\end{document}